\documentclass[a4paper,11pt]{article}
\usepackage{amsfonts}
\usepackage{amsmath}
\usepackage{amssymb}
\usepackage{array,booktabs,makecell}
\usepackage{mathtools}
\usepackage{colortbl}
\usepackage{tikz}
\usepackage{pgfplots}
\usepackage[flushleft]{threeparttable}
\usepackage{cite}
\usepackage{hyperref}
\usepackage{graphicx}
\usepackage{subcaption}
\numberwithin{equation}{section}
\usepackage[left=2cm,right=2cm,top=3.5cm,bottom=3.5cm]{geometry}

\allowdisplaybreaks[1]
\hypersetup{
	colorlinks=true,
	linkcolor=ceruleanblue,
	filecolor=ceruleanblue,      
	urlcolor=ceruleanblue,
	citecolor=ceruleanblue,
}
\definecolor{ceruleanblue}{rgb}{0.0, 0.2, 0.6}

\date{\today}

\begin{document}
	
	\begin{flushright} {\footnotesize YITP-23-45, IPMU23-0007, RUP-23-8}  \end{flushright}
	
	\begin{center}
		\LARGE{\bf Quasinormal Modes from EFT of Black Hole Perturbations with Timelike Scalar Profile}
		\\[1cm] 
		
		\large{Shinji Mukohyama$^{\,\rm a, \rm b}$, Kazufumi Takahashi$^{\,\rm a}$, Keitaro Tomikawa$^{\,\rm c}$ \\ and Vicharit Yingcharoenrat$^{\,\rm b}$}
		\\[0.5cm]
		
		\small{
			\textit{$^{\rm a}$
				Center for Gravitational Physics and Quantum Information, Yukawa Institute for Theoretical Physics, 
				\\ Kyoto University, 606-8502, Kyoto, Japan}}
		\vspace{.2cm}
		
		\small{
			\textit{$^{\rm b}$
				Kavli Institute for the Physics and Mathematics of the Universe (WPI), The University of Tokyo Institutes for Advanced Study (UTIAS), The University of Tokyo, Kashiwa, Chiba 277-8583, Japan}}
		\vspace{.2cm}
		
		\small{
			\textit{$^{\rm c}$
				Department of Physics, Rikkyo University, Toshima, Tokyo 171-8501, Japan}}
		
	\end{center}
	
	\vspace{0.3cm} 
	
	\begin{abstract}\normalsize
The Effective Field Theory (EFT) of perturbations on an arbitrary background geometry with a timelike scalar profile was recently constructed in the context of scalar-tensor theories. In this paper, we use this EFT to study quasinormal frequencies of odd-parity perturbations on a static and spherically symmetric black hole background. Keeping a set of operators that can accommodate shift-symmetric quadratic higher-order scalar-tensor theories, we demonstrate the computation for two examples of hairy black holes, of which one is the stealth Schwarzschild solution and the other is the Hayward metric accompanied by a non-trivial scalar field. We emphasize that this is the first phenomenological application of the EFT, opening a new possibility to test general relativity and modified gravity theories in the strong gravity regime. 
	\end{abstract}
	
	\vspace{0.3cm} 
	
	\vspace{2cm}
	
	\newpage
	{
		\hypersetup{linkcolor=black}
		\tableofcontents
	}
	
	\flushbottom
	
	\vspace{1cm}
	
	
\section{Introduction}
The direct detections of gravitational waves by LIGO and Virgo have opened up a new era of testing gravitational theories at strong-field/dynamical regimes~\cite{LIGOScientific:2016aoc,LIGOScientific:2018mvr,LIGOScientific:2020ibl,LIGOScientific:2021usb,LIGOScientific:2021djp}. In particular, the signal from the \emph{ringdown} phase of a merger of compact objects, such as black holes and neutron stars, has been considered to be a promising observational target. In general relativity (GR), it is well known that the gravitational wave signal at late times (ringdown) can be well studied by black hole perturbation theory~\cite{Regge:1957td,Zerilli:1970wzz,Teukolsky:1973ha}, which leads to the fact that it is well approximated by a superposition of the quasinormal modes (QNMs)~\cite{Buonanno:2006ui}. Each QNM is described by a specific complex frequency, called quasinormal frequency (QNF), whose real and imaginary parts correspond to the frequency of the temporal oscillation and the exponential damping rate, respectively. In the case of GR, the QNFs of a black hole depend only on its mass, electric charge, and angular momentum, thanks to the no-hair theorem~\cite{Bekenstein:1971hc,Bekenstein:1995un}. On the contrary, in modified gravity where a black hole may develop some non-trivial hair, the QNFs would potentially get affected by such an effect. In other words, the modification of gravity in the vicinity of compact objects can be encoded in the spectrum of QNMs. Therefore, the observation of the ringdown signal and the extraction of the QNFs would provide a hint of an underlying gravitational theory. 

In general, modified gravity theories tend to involve some additional propagating degrees of freedom. Scalar-tensor theories, which involve a single scalar field on top of the metric, have been extensively studied in many contexts such as cosmology and black hole physics until today. The most general scalar-tensor theory with second-order Euler-Lagrange equations in four spacetime dimensions is Horndeski's theory~\cite{Horndeski:1974wa,Deffayet:2011gz,Kobayashi:2011nu}, which offers a unifying description of traditional scalar-tensor theories, e.g., Brans-Dicke~\cite{Brans:1961sx} and k-essence theories~\cite{ArmendarizPicon:1999rj}. Note that the second-order nature of the Euler-Lagrange equations ensures the absence of the so-called Ostrogradsky ghost \cite{Woodard:2015zca}. Further development on the degenerate higher-order scalar-tensor (DHOST) theory~\cite{Langlois:2015cwa,Crisostomi:2016czh,BenAchour:2016fzp} was proposed as a generalization of Horndeski's theory by allowing the Euler-Lagrange equations to have higher-order derivatives satisfying the degeneracy condition~\cite{Motohashi:2014opa,Langlois:2015cwa,Motohashi:2016ftl,Klein:2016aiq}. A further generalization is the so-called U-DHOST theory~\cite{DeFelice:2018mkq,DeFelice:2021hps,DeFelice:2022xvq}, in which the degeneracy condition is satisfied only in the unitary gauge. U-DHOST theories in general have a non-propagating shadowy mode that satisfies an elliptic differential equation on a spacelike hypersurface~\cite{DeFelice:2018mkq,DeFelice:2021hps}. Yet further generalizations were achieved by performing a higher-derivative generalization of invertible disformal transformation~\cite{Takahashi:2021ttd} on the Horndeski and U-DHOST theories, resulting in the generalized disformal Horndeski (GDH)~\cite{Takahashi:2022mew} and generalized disformal unitary-degenerate (GDU) theories~\cite{Takahashi:2023jro}, respectively.\footnote{It should be noted that an invertible transformation does not change the number of physical degrees of freedom~\cite{Domenech:2015tca,Takahashi:2017zgr}. A non-invertible disformal transformation can also be used to generate a class of ghost-free theories, but the resultant theories do not accommodate viable cosmology~\cite{Takahashi:2017pje,Langlois:2018jdg}.} The generalized disformal transformation could introduce Ostrogradsky modes when matter fields are taken into account, but there exists a non-trivial class where matter fields can be consistently coupled~\cite{Takahashi:2022mew,Naruko:2022vuh,Takahashi:2022ctx,Ikeda:2023ntu}. The GDH/GDU theories form the most general class of healthy covariant scalar-tensor theories known so far. All of the scalar-tensor theories mentioned above and more general ones beyond them should be universally described by an effective field theory (EFT) given that the symmetry breaking pattern and the background geometry are specified. 

In general, the EFT provides a model-independent framework to study dynamics of perturbations on a particular background. The essential ingredients of the EFT are the field contents, the symmetry breaking pattern, and the background spacetime. An EFT of scalar-tensor theories on Minkowski and de Sitter spacetimes, dubbed ghost condensation, was formulated in \cite{Arkani-Hamed:2003pdi,Arkani-Hamed:2003juy} assuming that a timelike gradient of a scalar field spontaneously breaks the time diffeomorphism and that the EFT remains invariant under the spatial diffeomorphism, the time translation and the time reflection up to the shift and the reflection of the scalar field. This formulation of the EFT action was later extended to a general Friedmann-Lema\^{\i}tre-Robertson-Walker background in \cite{Cheung:2007st,Gubitosi:2012hu}, called the EFT of inflation/dark energy. Recently, in \cite{Mukohyama:2022enj}, the EFT action with a timelike scalar profile was generalized to an arbitrary background geometry.\footnote{See a similar construction of the EFT that applies to the shift-symmetric Horndeski theories.} The dictionary has been developed to relate such an EFT to concrete covariant theories such as Horndeski's theory~\cite{Mukohyama:2022enj} and shift-symmetric quadratic higher-order scalar-tensor (HOST) theories~\cite{Mukohyama:2022skk}. In fact, thanks to the timelike nature of a scalar field, the EFT of \cite{Mukohyama:2022enj} can describe physics at both cosmological and black hole scales in a single framework. With this reasoning, it offers a possibility to extract some information about scalar-field dark energy from observations associated with astrophysical black holes, at least in principle. This fact adds another dimension of interest to the EFT and motivates us to call it the EFT of black hole perturbations with a timelike scalar profile. Note that an EFT of black hole perturbations with a spacelike scalar profile on a static and spherically symmetric background was constructed in \cite{Franciolini:2018uyq} (see also \cite{Hui:2021cpm} for the formulation of the EFT on a slowly rotating black hole).

The main purpose of this paper is to study the spectrum of QNMs using the EFT of perturbations on a static and spherically symmetric black hole with a timelike scalar profile.\footnote{In our EFT or any modified gravity theories, the way how matter fields excite QNMs would be different from GR in general. Having said that, the computation of QNFs is an important step towards testing modified gravity theories with the ringdown signal.} The linear perturbations about such a spherically symmetric spacetime can be decomposed into odd- and even-parity modes, and they evolve independently. The master equation for the odd modes based on the EFT was derived in \cite{Mukohyama:2022skk} as a generalization of the Regge-Wheeler (RW) equation in GR. In the present paper, we reformulate the generalized RW equation in a more ready-to-use form and compute the QNFs for the odd modes. For demonstration purposes, we consider two examples of hairy black holes: One is the stealth Schwarzschild solution and the other is the Hayward metric with a non-trivial scalar field. We will clarify how the modification from the GR solution, i.e., the Schwarzschild solution, affects the QNFs. 

This paper is organized as follows. In Section~\ref{sec:setup}, we give a brief review of the EFT of perturbations on an arbitrary background with a timelike scalar profile~\cite{Mukohyama:2022enj} and revisit the analysis of odd-parity perturbations on a static and spherically symmetric background based on \cite{Mukohyama:2022skk}. In Section~\ref{sec:QNM}, we demonstrate the computation of the QNFs with the two examples of hairy black hole solutions mentioned above. Finally, we draw our conclusions in Section~\ref{sec:conclusions}.

\section{Setup}\label{sec:setup}
\subsection{Review of the EFT}\label{sec:EFT_review}
In this Subsection, we give a brief review of the EFT of perturbations on an arbitrary background with a timelike scalar profile~\cite{Mukohyama:2022enj,Mukohyama:2022skk}. See also \cite{Khoury:2022zor} for a similar construction of the EFT for shift-symmetric scalar-tensor theories. The main idea of our EFT, similar to the EFT of ghost condensation~\cite{Arkani-Hamed:2003pdi} and the EFT of inflation/dark energy~\cite{Cheung:2007st,Gubitosi:2012hu}, is that the time-dependent background of the scalar field, $\bar{\Phi}$, spontaneously breaks the time diffeomorphism and defines a preferred ($\Phi = const.$) time-slicing whose unit normal vector can be defined by 
\begin{align}\label{eq:normal_EFT}
n_\mu \equiv 
-\frac{\partial_\mu \Phi}{\sqrt{-X}}
\rightarrow - \frac{\delta_\mu^\tau}{\sqrt{-g^{\tau\tau}}} \;,
\end{align}
with $X\equiv g^{\mu\nu}\partial_\mu\Phi\partial_\nu\Phi$ being the kinetic term of the scalar field, so that $n_\mu n^\mu = -1$. Here, we use $\tau$ as the time coordinate such that $\bar{\Phi}=\bar{\Phi}(\tau)$ and $\delta\Phi\equiv\Phi-\bar{\Phi}=0$ (unitary gauge). The expression to the right of the arrow in (\ref{eq:normal_EFT}) refers to the one in the unitary gauge. Then, the residual symmetry of the EFT is the 3d diffeomorphism invariance. Therefore, the EFT we are going to write down in the unitary gauge can contain a scalar function of, for example, the 4d and the 3d curvatures, the extrinsic curvature, the ($\tau\tau$)-component of the inverse metric tensor, and the time coordinate~$\tau$. 

Let us, for convenience, introduce the Arnowitt-Deser-Misner (ADM) $3+1$ decomposition, where the metric can be written in the following form:
\begin{align}
ds^2 = -N^2 d\tau^2 + h_{ij} (dx^i + N^i d\tau) (dx^j + N^j d\tau) \;, 
\end{align}
where $N$ is the lapse function, $N^i$ is the shift vector, and $h_{\mu\nu} \equiv g_{\mu\nu} + n_\mu n_\nu$ is the induced metric on a spacelike hypersurface of constant $\tau$. The indices~$i,j$ denote the spatial components. Note that the spatial indices are raised and lowered by the induced metric~$h_{ij}$.
 
As usual, we define the extrinsic curvature using the induced metric~$h_{\mu\nu}$ as
 \begin{align}
 K_{\mu\nu} \equiv h_\mu^\rho \nabla_\rho n_\nu \;,
 \end{align}
where $\nabla_\mu$ is the 4d covariant derivative. In particular, one obtains the spatial components of $K_{\mu\nu}$ and its trace in terms of the ADM variables:
\begin{align}
K_{ij} = \frac{1}{2N}\left(\dot{h}_{ij} - D_i N_j - D_j N_i\right) \;, \qquad K = h^{ij} K_{ij} \;,
\end{align}
with a dot being the derivative with respect to $\tau$ and $D_i$ the 3d covariant derivative constructed from the induced metric~$h_{ij}$. In addition, the 3d curvature~${}^{(3)}\!R$ can be computed in a usual way using the induced metric~$h_{ij}$.

As discussed earlier, the EFT action we are going to write down in the unitary gauge is invariant under the 3d diffeomorphism. Therefore, apart from the 4d covariant terms such as the 4d Ricci scalar~$\tilde{R}$, the action can depend on any geometrical quantities that are covariant under the 3d diffeomorphism, for instance $g^{\tau\tau}$ ($=-1/N^2$), $K_{\mu\nu}$, and ${}^{(3)}\!R_{\mu\nu}$. Note that the symbol~$R$ (without the tilde) is reserved to denote the 4d Ricci scalar with the divergence term subtracted [see Eq.~(\ref{eq:4d_Ricci})]. Moreover, the action can depend on $\tau$ explicitly. With all the possibilities above, the unitary-gauge action takes the form 
\begin{align}\label{eq:action_uni}
S = \int d^4x \sqrt{-g}~F(\tilde{R}_{\mu\nu\alpha\beta}, K_{\mu\nu}, g^{\tau\tau}, \nabla_\mu, \tau) \;,
\end{align}
where $F$ is a scalar function of those 4d and 3d diffeomorphism covariant quantities, and $\tilde{R}_{\mu\nu\alpha\beta}$ is the 4d Riemann tensor of the metric~$g_{\mu\nu}$. Note that the 3d curvature tensor can be written in terms of the 4d curvature and the extrinsic curvature by use of the Gauss relation, so that we did not include it explicitly in the action. It is important to point out that the action~(\ref{eq:action_uni}) can be applied to any background geometries without assuming any symmetries of the background. However, this form of the action is not the one we are going to use for practical calculations. In practice, we expand it up to the necessary order in perturbations and derivatives. Then, as we will see below, one needs to impose a set of consistency relations among the expansion coefficients (i.e., EFT parameters) so that the resulting EFT action respects the 3d diffeomorphism invariance as a whole. After imposing the consistency conditions, one can consider the remaining independent coefficients as arbitrary functions. 

Following \cite{Mukohyama:2022enj}, we define the perturbations as follows:
\begin{align}
\delta g^{\tau\tau} \equiv g^{\tau\tau} - \bar{g}^{\tau\tau}(\tau, \vec{x}) \;, \qquad \delta K^\mu_\nu \equiv K^\mu_\nu - \bar{K}^\mu_\nu(\tau, \vec{x}) \;, \qquad \delta {}^{(3)}\!R^\mu_\nu \equiv {}^{(3)}\!R^\mu_\nu -  {}^{(3)}\!\bar{R}^\mu_\nu(\tau, \vec{x}) \;, \label{eq:pert}
\end{align}
where a bar denotes the background value. Practically, the EFT action is written as a polynomial of these perturbation variables as well as their derivatives. Each EFT coefficient can have an explicit dependence on both $\tau$ and $\vec{x}$ due to the spacetime dependence of the background quantities~$\bar{g}^{\tau\tau}$, $\bar{K}^\mu_\nu$, and ${}^{(3)}\!\bar{R}^\mu_\nu$. Note that such a dependence on $\vec{x}$ is not compatible with the 3d diffeomorphism invariance in general. Therefore, in order for the EFT action to respect the 3d diffeomorphism invariance, we impose a set of consistency relations on the EFT coefficients. As explained in great detail in \cite{Mukohyama:2022enj}, such  non-trivial consistency relations can be obtained by applying the chain rule with respect to the spatial derivative to each term of the Taylor expansion of the action~(\ref{eq:action_uni}). For example, one obtains 
\begin{align}
\frac{\partial }{\partial x^i}\bar{F} = \bar{F}_{g^{\tau\tau}} \frac{\partial \bar{g}^{\tau\tau}}{\partial x^i} + \bar{F}_K \frac{\partial \bar{K}}{\partial x^i} + \bar{F}_{\sigma^\mu_\nu} \frac{\partial \bar{\sigma}^\mu_\nu}{\partial x^i} + \bar{F}_{{}^{(3)}\!R} \frac{\partial {}^{(3)}\!\bar{R}}{\partial x^i} + \bar{F}_{r^\mu_\nu} \frac{\partial \bar{r}^\mu_\nu}{\partial x^i}\;,
\end{align}
where $\sigma_{\mu\nu} \equiv K_{\mu\nu} - h_{\mu\nu} K/3$, $r_{\mu\nu} \equiv {}^{(3)}\!R_{\mu\nu} - h_{\mu\nu} {}^{(3)}\!R/3$, and we have omitted terms that are higher order in derivatives. The notation~$\bar{F}_X = (\partial F/\partial X)_{\rm BG}$ refers to the Taylor coefficient evaluated on the background. This relation tells us that the Taylor coefficients are not independent of each other, but they are related in a non-trivial way such that at each order in perturbations one recovers the 3d diffeomorphism invariance of the action. In general, one obtains infinitely many consistency relations which are coming from applying the chain rule to other coefficients of the Taylor expansion. Since the action is allowed to explicitly depend on $\tau$, the relations derived from the chain rule with respect to the $\tau$-derivative are trivially satisfied. Therefore, we conclude that imposing the set of consistency relations associated with $\vec{x}$-derivatives guarantees that the EFT action is invariant under the 3d diffeomorphism. However, if we restrict ourselves to shift-symmetric scalar-tensor theories, the chain rule with respect to the $\tau$-derivatives would result in an additional set of infinitely many constraints on the EFT coefficients \cite{Khoury:2022zor}.\footnote{The constraints in the shift-symmetric EFT of inflation were obtained in \cite{Finelli:2018upr}. Further development was studied in the context of  EFT of vector-tensor theories~\cite{Aoki:2021wew} constructed on a cosmological background, where the shift symmetry become gauged~\cite{Cheng:2006us} and one is required to impose a set of consistency relations associated with the time coordinate such that the action remains invariant under the combined $U(1)$ and time diffeomorphism.} 

Now we are ready to write down the EFT action. For simplicity, as we did in \cite{Mukohyama:2022skk}, we omit terms containing $r_{\mu\nu}$ as we are mostly interested in a minimal EFT action that accommodates the shift-symmetric quadratic HOST theories. We therefore have 
\begin{align}\label{eq:EFT_shift_Z2}
		S = \int d^4x \sqrt{-g} \bigg[&\frac{M_\star^2}{2}
		R - \Lambda(y) - c(y)g^{\tau\tau} - \tilde{\beta}(y) K - \alpha(y)\bar{K}^\mu_\nu K^{\nu}_\mu -\zeta(y) n^\mu\partial_\mu g^{\tau\tau} \nonumber \\ 
		& + \frac{1}{2} m_2^4(y) (\delta g^{\tau\tau})^2 + \frac{1}{2} \tilde{M}_1^3(y) \delta g^{\tau\tau} \delta K + \frac{1}{2} M_2^2(y) \delta K^2 + \frac{1}{2} M_3^2(y) \delta K^\mu_\nu \delta K^\nu_\mu \nonumber \\
		& + \frac{1}{2}\mu_1^2(y) \delta g^{\tau\tau} \delta {}^{(3)}\!R + \frac{1}{2} \lambda_1(y)^\mu_\nu \delta g^{\tau\tau} \delta K^\nu_\mu + \frac{1}{2}{\mathcal M}_1^2(y)(\bar{n}^\mu\partial_\mu\delta g^{\tau\tau})^2 \nonumber \\
		& +\frac{1}{2}{\mathcal M}_2^2(y)\delta K(\bar{n}^\mu\partial_\mu\delta g^{\tau\tau})+\frac{1}{2}{\mathcal M}_3^2(y)\bar{h}^{\mu\nu}\partial_\mu\delta g^{\tau\tau}\partial_\nu\delta g^{\tau\tau}+\cdots \bigg] \;,
	\end{align}
where $y = \{\tau, \vec{x}\}$ and we have defined
    \begin{align}\label{eq:4d_Ricci}
    R \equiv {}^{(3)}\!R + K_{\mu\nu}K^{\mu\nu} - K^2 = \tilde{R} - 2\nabla_\mu( K n^\mu - n^\nu\nabla_\nu n^\mu)\;.
    \end{align}
Note that at this stage we have not assumed any specific forms of both $\bar{\Phi}(\tau)$ and $\bar{g}_{\mu\nu}$. 
Here, we work in the Einstein frame where the coefficient in front of $R$ is a constant. As we will see in the next Subsection, the first six terms in (\ref{eq:EFT_shift_Z2}) contribute to the background equations of motion, giving rise to the tadpole cancellation conditions. Also, the coefficients should satisfy the consistency conditions mentioned above. It should be noted that this EFT action accommodates Horndeski theories as well as quadratic HOST theories. Indeed, the dictionary with the Horndeski theory was reported in \cite{Mukohyama:2022enj}, while the one with the shift- and $Z_2$-symmetric subclass of HOST theories was presented in \cite{Mukohyama:2022skk}.\footnote{Likewise, it would also be straightforward to extend the EFT to accommodate GDH/GDU theories (see \cite{Takahashi:2023jro} for the case of EFT of cosmological perturbations).} In the latter paper, we have imposed $\bar{\Phi}\propto \tau$ and $\bar{g}^{\tau\tau}=-1$ (so that $\bar{X}=const.$) for simplicity.

In the next Subsection, we will express the tadpole cancellation conditions derived from the EFT action (\ref{eq:EFT_shift_Z2}) on a static and spherically symmetric background metric.

\subsection{Static and spherically symmetric background}\label{sec:background}
As discussed in the previous Subsection, our EFT can be used to study the dynamics of perturbations on generic background geometries. In this paper, as a first step towards tests of gravity against observational data, we would like to use our EFT to determine the spectrum of QNMs on a static and spherically symmetric background. In particular, we are going to use the EFT that accommodates shift-symmetric quadratic HOST theories. Below, we will review the background equations of motion derived from our EFT. 

First, let us consider the following static and spherically symmetric metric:
\begin{align}\label{eq:metric_BG}
	\bar{g}_{\mu\nu}dx^\mu dx^\nu = -A(r)dt^2 + \frac{dr^2}{B(r)} + r^2 d\Omega^2 \;,
	\end{align}
where $d\Omega^2 = d\theta^2 + \sin^2\theta\,d\phi^2$ and $A(r)$, $B(r)$ are functions of $r$. Written in terms of the so-called Lema\^{\i}tre coordinates~\cite{Lemaitre:1933gd,Khoury:2020aya}, we have
	\begin{align}\label{eq:BG_Lemaitre}
	\bar{g}_{\mu\nu}dx^\mu dx^\nu = -d\tau^2 + [1 - A(r)] d\rho^2 + r^2 d\Omega^2 \;,
	\end{align}
where the coordinates~$\tau$ and $\rho$ are respectively related to $t$ and $r$ via 
	\begin{align}
	d\tau = dt + \sqrt{\frac{1 - A}{AB}}\,dr \;, \qquad d\rho = dt + \frac{dr}{\sqrt{AB(1 - A)}} \;. \label{eq:trans_Lemai}
	\end{align}
Moreover, it is useful to note that $r$ is a function of $\rho - \tau$, which satisfies
\begin{align}
\partial_\rho r = - \dot{r} = \sqrt{\frac{B(1 - A)}{A}} \;,
\end{align}
with a dot denoting the derivative with respect to $\tau$.

Let us now proceed to the background equations of motion. We recall that we are now interested in static and spherically symmetric background in shift-symmetric theories. In fact, the staticity of the background metric requires that the background scalar field has to be linear in $\tau$, giving rise to $\bar{X} = const$. Therefore, the EFT action~(\ref{eq:EFT_shift_Z2}) becomes
\begin{align}
		S = \int d^4x \sqrt{-g} \bigg[&\frac{M_\star^2}{2}R - \Lambda(r) - c(r)g^{\tau\tau} - \tilde{\beta}(r) K - \alpha(r)\bar{K}^\mu_\nu K^{\nu}_\mu -\zeta(r) n^\mu\partial_\mu g^{\tau\tau} \nonumber \\ 
		& + \frac{1}{2} m_2^4(r) (\delta g^{\tau\tau})^2 + \frac{1}{2} \tilde{M}_1^3(r) \delta g^{\tau\tau} \delta K + \frac{1}{2} M_2^2(r) \delta K^2 + \frac{1}{2} M_3^2(r) \delta K^\mu_\nu \delta K^\nu_\mu \nonumber \\
		& + \frac{1}{2}\mu_1^2(r) \delta g^{\tau\tau} \delta {}^{(3)}\!R + \frac{1}{2} \lambda_1(r)^\mu_\nu \delta g^{\tau\tau} \delta K^\nu_\mu + \frac{1}{2}{\mathcal M}_1^2(r)(\bar{n}^\mu\partial_\mu\delta g^{\tau\tau})^2 \nonumber \\
		& +\frac{1}{2}{\mathcal M}_2^2(r)\delta K(\bar{n}^\mu\partial_\mu\delta g^{\tau\tau})+\frac{1}{2}{\mathcal M}_3^2(r)\bar{h}^{\mu\nu}\partial_\mu\delta g^{\tau\tau}\partial_\nu\delta g^{\tau\tau}+\cdots \bigg] \;,
		\label{eq:EFT_HOST}
	\end{align}
where the EFT coefficients are now functions of $r$ only. Then, the background equations of motion (tadpole cancellation conditions) read
    \begin{equation}
	\begin{split}
	\Lambda-c
	&=M_\star^2(\bar{G}^\tau{}_\rho-\bar{G}^\rho{}_\rho)\;, \\
	\Lambda+c+\frac{2}{r^2}\sqrt{\frac{B}{A}}\left(r^2\sqrt{1-A}\,\zeta\right)'
	&=-M_\star^2\bar{G}^\tau{}_\tau\;, \\
	\left[\partial_\rho\bar{K}+\frac{1-A}{r}\left(\frac{B}{A}\right)'\,\right]\alpha+\frac{A'B}{2A}\alpha'+\sqrt{\frac{B(1 - A)}{A}}\tilde{\beta}'
	&=-M_\star^2\bar{G}^\tau{}_\rho\;, \\
	\frac{1}{2r^2}\sqrt{\frac{B}{A}}\left[r^4\sqrt{\frac{B}{A}}\left(\frac{1-A}{r^2}\right)'\alpha\right]'
	&=M_\star^2(\bar{G}^\rho{}_\rho-\bar{G}^\theta{}_\theta)\;,
	\end{split} \label{EOM_BG}
    \end{equation}
with the relevant components of the background Einstein tensor given by
    \begin{equation}
	\begin{split}
	\bar{G}^\tau{}_\tau&=-\frac{[r(1-B)]'}{r^2}+\frac{1-A}{r}\left(\frac{B}{A}\right)'\;, \\
	\bar{G}^\tau{}_\rho&=-\frac{1-A}{r}\left(\frac{B}{A}\right)'\;, \\
	\bar{G}^\rho{}_\rho&=-\frac{[r(1-B)]'}{r^2}-\frac{1}{r}\left(\frac{B}{A}\right)'\;, \\
	\bar{G}^\theta{}_\theta&=\frac{B(r^2A')'}{2r^2A}+\frac{(r^2A)'}{4r^2}\left(\frac{B}{A}\right)'\;,
	\end{split}
    \end{equation}
where a prime denotes the derivative with respect to $r$. Note that the background equations above receive contributions only from the terms in the first line of (\ref{eq:EFT_HOST}).\footnote{If one is interested in, e.g., shift-symmetric cubic HOST theories, one then has to take into account the term linear in $r_{\mu\nu}$, which would contribute to the background equations~(\ref{EOM_BG}) as well.} These equations provide relations among the EFT coefficients through given functions~$A(r)$ and $B(r)$.

Before closing this Subsection, let us consider the case where $A(r) = B(r)$. This indeed will be our main focus in Section~\ref{sec:QNM}, in which we compute QNFs. Assuming that $A(r) = B(r)$, the fourth equation in \eqref{EOM_BG} yields 
    \begin{equation}
    r^4\left(\frac{1-A}{r^2}\right)'(\alpha-M_\star^2)= const.\;,
    \end{equation}
which restricts $\alpha$ to be of the form 
    \begin{equation}\label{eq:alpha2Mstar}
    \alpha=M_\star^2+\frac{3\lambda}{r(2-2A+rA')}\;,
    \end{equation}
with $\lambda$ being a constant.

In the next Subsection, we will study the dynamics of odd-parity perturbations and provide the master equation necessary to compute the QNM spectrum.

	
\subsection{Odd-parity perturbations}\label{sec:odd-parity}
In the previous Subsection, we obtained the background equations of motion for a static and spherically symmetric background derived from the EFT action~(\ref{eq:EFT_HOST}). Here, we revisit the analysis of linear odd-parity perturbations (also called axial perturbations) with general higher multipoles~$\ell \geq 2$ around the background metric~(\ref{eq:BG_Lemaitre}).\footnote{In the odd-parity sector, the dipole modes with $\ell=1$ do not propagate and is related to the slow rotation of a black hole. We refer the readers to Section~5.2 of \cite{Mukohyama:2022skk} for details.} Most of the results in this Section were already presented in \cite{Mukohyama:2022skk}, but here we reformulate the master equation in a form ready for practical applications.

Let us first introduce the notation we are using for metric perturbations. Since the background metric~(\ref{eq:metric_BG}) has the $SO(2)$ symmetry, it is convenient to decompose the perturbations into the odd and even sectors on the basis of spherical harmonics. Following the notation in, for instance \cite{Takahashi:2021bml,Mukohyama:2022skk}, we introduce the perturbations, $\delta g_{\mu\nu} \equiv g_{\mu\nu} - \bar{g}_{\mu\nu}$, in the odd sector as follows:
\begin{align}
\begin{split}
\delta g_{\tau\tau} &= \delta g_{\tau\rho} = \delta g_{\rho\rho} = 0 \;, \\
\delta g_{\tau a} &= \sum_{\ell,m} r^2 h_{0,\ell m}(\tau,\rho) E_a^{\ b} \bar{\nabla}_b Y_{\ell m}(\theta, \phi) \;,  \\
\delta g_{\rho a} &= \sum_{\ell,m} r^2 h_{1,\ell m}(\tau,\rho) E_a^{\ b} \bar{\nabla}_b Y_{\ell m}(\theta, \phi) \;, \\
\delta g_{ab} &=  \sum_{\ell,m} r^2 h_{2,\ell m}(\tau,\rho) E_{(a|}^{\ \ \ c} \bar{\nabla}_c \bar{\nabla}_{|b)} Y_{\ell m}(\theta, \phi)  \;,
\end{split}
\end{align}
where $Y_{\ell m}$ is the spherical harmonics, $E_{ab}$ is the completely anti-symmetric rank-2 tensor defined on a $2$-sphere, $\bar{\nabla}_a$ is the covariant derivative defined with the unit $2$-sphere metric, and the indices~$a, b, \cdots$ denote the angular variables~$\{\theta, \phi\}$ . Under an infinitesimal coordinate transformation~$x^\mu \rightarrow x^\mu + \epsilon^\mu$ with 
\begin{align}
\epsilon^\tau = \epsilon^\rho = 0 \;, \qquad \epsilon^a = \sum_{\ell,m} \Xi_{\ell m}(\tau,\rho) E^{ab} \bar{\nabla}_b Y_{\ell m}(\theta,\phi) \;,
\end{align}
the perturbations~$h_0$, $h_1$, and $h_2$ are transformed as
\begin{align}
h_0 \rightarrow h_0 - \dot{\Xi} \;, \qquad h_1 \rightarrow h_1 - \partial_\rho \Xi \;, \qquad h_2 \rightarrow h_2 - 2 \Xi \;.
\end{align}
We see that choosing $\Xi = h_2/2$ completely fixes the gauge such that $h_2 \rightarrow 0$. Note that this gauge fixing can be done at the Lagrangian level for the modes with $\ell \geq 2$~\cite{Motohashi:2016prk}. Also, without loss of generality, we will only focus on the modes with $m = 0$ due to the spherical symmetry of the background, so that the analysis only involves the Legendre polynomials~$P_\ell(\cos\theta)$ instead of the spherical harmonics. In what follows, the perturbation variables will refer to the coefficients of $P_{\ell}(\cos\theta)$.

Let us now consider the EFT action in the odd sector. We first note that any quantities evaluated on the background metric are even under a parity transformation. Also, the EFT operators that contain, for example, $\delta g^{\tau\tau}$ and $\delta K$ do not contribute to the odd sector. Therefore, the EFT action in the odd-parity sector up to second order in perturbations reads 
	\begin{align}
	S_{\rm odd} = \int d^4x \sqrt{-g} \bigg[\frac{M_\star^2}{2}R - \Lambda(r) - c(r)g^{\tau\tau} -\tilde{\beta}(r) K - \alpha(r)\bar{K}^{\mu}_\nu K^\nu_{\mu} + \frac{1}{2} M_3^2(r) \delta K^\mu_\nu \delta K^\nu_\mu \bigg] \;. \label{eq:EFT_action}
	\end{align}
In what follows, we outline the procedure to obtain the quadratic Lagrangian for the physical degree of freedom in the odd sector based on \cite{Mukohyama:2022skk}. The quadratic Lagrangian for $h_0$ and $h_1$ from the action~(\ref{eq:EFT_action}) can be written in the form 
\begin{align}
		\frac{2\ell + 1}{2 \pi j^2}\mathcal{L}_2 = p_1 h_0^2 + p_2 h_1^2 + p_3 [(\dot{h}_1 - \partial_\rho h_0)^2 + 2p_4 h_1 \partial_\rho h_0] \;, \label{L2_odd_Ein}
\end{align}
where $j^2 \equiv \ell(\ell + 1)$ and the parameters $p$'s are defined in terms of the EFT coefficients by
\begin{equation}
	\begin{split}
		p_1 &\equiv  \frac{1}{2}(j^2-2)r^2\sqrt{1-A}\,(M_\star^2 + M_3^2) \;, \qquad
		p_2 \equiv -(j^2-2)\frac{r^2M_\star^2}{2\sqrt{1-A}} + (p_3p_4)^{\boldsymbol{\cdot}} \;, \\
		p_3 &\equiv \frac{(M^2_\star + M_3^2) r^4}{2\sqrt{1 - A}} \;, \qquad
		p_4 \equiv \sqrt{\frac{B}{A(1 - A)}}\left(\frac{A'}{2}+\frac{1-A}{r}\right) \frac{\alpha + M_3^2}{M_\star^2 + M_3^2} \;.
	\end{split} \label{eq:odd_2_ein}
\end{equation}
Note that the parameter~$p_4$ is vanishing when the condition~$\alpha + M_3^2 = 0$ is satisfied, which is realized, e.g., in shift-symmetric HOST theories~\cite{Takahashi:2019oxz,Tomikawa:2021pca,Mukohyama:2022skk}. In this sense, the parameter~$p_4$ is peculiar to our EFT. However, as clarified in \cite{Mukohyama:2022skk}, a non-vanishing $p_4$ leads to the absence of a slowly rotating black hole solution or otherwise the divergence of the radial sound speed at spatial infinity. Therefore, in what follows, we restrict ourselves to the case where $p_4 = 0$, i.e., $\alpha+M_3^2=0$.

Let us rewrite the Lagrangian~\eqref{L2_odd_Ein} with $p_4=0$ in terms of a master variable, which represents the single propagating degree of freedom in the odd sector. For this purpose, we introduce an auxiliary field~$\chi$ as follows:
\begin{align}
	\frac{2\ell + 1}{2 \pi j^2}\mathcal{L}_2 = p_1 h_0^2 + p_2 h_1^2
	+ p_3[-\chi^2 + 2\chi (\dot{h}_1 - \partial_\rho h_0)] \;.
\end{align}
One can recover the Lagrangian~\eqref{L2_odd_Ein} by integrating out $\chi$. On the other hand, integrating out $h_0$ and $h_1$ from the above Lagrangian, one obtains the following expression: 
\begin{align}\label{eq:chi_s}
	\frac{(j^2 - 2)(2\ell + 1)}{2 \pi j^2} \mathcal{L}_2 = s_1 \dot{\chi}^2 - s_2 (\partial_\rho\chi)^2 - s_3 \chi^2 \;,
\end{align}
with the parameters $s$'s defined as 
\begin{equation}
	s_1 \equiv  -\frac{(j^2 - 2)p_3^2}{p_2} \;, \qquad
	s_2 \equiv \frac{(j^2-2)p_3^2}{p_1} \;, \qquad
	s_3 \equiv  (j^2-2)p_3\left[1 - \left(\frac{p_1 + p_2}{p_1p_2}\dot{p}_3\right)^{\boldsymbol{\cdot}} \,\right] \;. \label{eq:para_s}
\end{equation}
Therefore, the variable~$\chi$ plays the role of the master variable and describes the propagating physical degree of freedom in the odd sector.\footnote{In HOST theories, there could be an Ostrogradsky mode associated with the higher derivatives of the scalar field in general, but it does not show up in the odd sector~\cite{Tomikawa:2021pca}. Also, even if the scalar field is nondynamical (which happens, e.g., in the cuscuton model~\cite{Afshordi:2006ad}, its extension~\cite{Iyonaga:2018vnu,Iyonaga:2020bmm}, or a related theory of type-\mbox{I\hspace{-1pt}I} minimally modified gravity~\cite{DeFelice:2020eju,DeFelice:2022uxv}), there remains one dynamical degree of freedom in the odd sector. The extra/missing degree of freedom is expected to be seen in the even sector.} From the Lagrangian~(\ref{eq:chi_s}), one obtains the radial and angular sound speeds for the odd modes as
    \begin{align}
    c_\rho^2 = \frac{\bar{g}_{\rho\rho}}{|\bar{g}_{\tau\tau}|}\frac{s_2}{s_1}\;, \qquad
    c_\theta^2 = \lim_{\ell\to\infty}\frac{r^2}{|\bar{g}_{\tau\tau}|}\frac{s_3}{j^2s_1}\;.
    \end{align}
Interestingly, under the condition~$\alpha+M_3^2=0$, the sound speeds for the odd modes in the radial and angular directions coincide, which we denote by $c_T^2$. Written explicitly in terms of EFT parameters, we have\footnote{Asymptotically far way from the source, the almost simultaneous detection of the GW event~GW170817 and the gamma-ray burst~170817A emitted from a binary neutron star merger puts the bound on $c_T^2$: $|1 - c_T^2| \lesssim 10^{-15}$ within LIGO/Virgo frequency band~\cite{TheLIGOScientific:2017qsa,GBM:2017lvd,Monitor:2017mdv}, ruling out a wide class of dark energy models in the context of scalar-tensor theories, see, e.g., \cite{Creminelli:2017sry,Sakstein:2017xjx,Ezquiaga:2017ekz}.}
    \begin{align}\label{eq:cT}
    \alpha_T(r)\equiv c_T^2-1
    =-\frac{M_3^2(r)}{M_\star^2+M_3^2(r)}\;.
    \end{align}
Note that the absence of ghost/gradient instabilities requires \cite{Mukohyama:2022skk}
    \begin{align}
    M_\star^2>0\;, \qquad
    M_\star^2+M_3^2>0\;,
    \end{align}
where the second condition can be equivalently written as $1+\alpha_T>0$.\footnote{It should be noted that this condition only applies at the linear level. The stability of perturbations in the presence of non-linearities was analyzed, for instance, in \cite{Creminelli:2019kjy} in the context of EFT of dark energy.} The fact that the odd-mode sound speed is different from that of light implies that the horizon for the odd modes is not the same as the one for photons. Note that these stability conditions are consistent with the ones obtained in \cite{Khoury:2020aya,Takahashi:2021bml}. 

For the purpose of computing QNFs, it is useful to go back to the static coordinate system. In terms of the Schwarzschild coordinates~$(t,r,\theta,\phi)$, the quadratic Lagrangian~\eqref{eq:chi_s} takes the form
\begin{align}\label{eq:L2_odd_a}
	\frac{(j^2 - 2)(2\ell + 1)}{2 \pi j^2} \mathcal{L}_2 = a_1 (\partial_t \chi)^2 - a_2 (\partial_r \chi)^2 + 2a_3 (\partial_t \chi) (\partial_r \chi) - a_4\chi^2 \;,
\end{align}
which involves a cross term~$(\partial_t\chi)(\partial_r\chi)$. The parameters $a$'s are given by
\begin{equation}
    \begin{split}
	&a_1 \equiv \frac{s_1-(1-A)^2s_2}{\sqrt{A^3B(1-A)}}\;, \qquad
	a_2 \equiv \sqrt{\frac{B(1-A)}{A}}(s_2-s_1)\;, \\
	&a_3 \equiv \frac{(1-A)s_2-s_1}{A}\;, \qquad
	a_4 \equiv \sqrt{\frac{A}{B(1-A)}}s_3\;.
	\end{split}
\end{equation}
Fortunately, the unwanted cross term in Eq.~(\ref{eq:L2_odd_a}) can be removed by the following redefinition of the time coordinate:
    \begin{align}\label{eq:tildet}
    t\quad\to\quad
	\tilde{t} = t + \int \frac{a_3}{a_2} dr
	=t+ \int \sqrt{\frac{1-A}{AB}}\frac{\alpha_T}{A+\alpha_T} dr\;.
\end{align}
Notice that generically $\alpha_T$ can be a non-trivial function of the radial coordinate, as we will discuss in Section~\ref{sec:Hayward}. Therefore, the quadratic Lagrangian can be recast in the following form:
\begin{align}\label{eq:Lagrangian_chi}
	\frac{(j^2 - 2)(2\ell + 1)}{2 \pi j^2} \mathcal{L}_2  = \tilde{a}_1 (\partial_{\tilde{t}} \chi)^2 - a_2 (\partial_r \chi)^2 - a_4\chi^2 \;,
\end{align}
with $\tilde{a}_1 \equiv a_1 + a_3^2/a_2$. We will discuss a possible subtlety arising from the use of coordinate~$\tilde{t}$ instead of the Killing time~$t$ in the next Section.

We are now ready to write down the master equation for the odd modes, i.e., the generalized RW equation. Let us define the generalized tortoise coordinate by\footnote{Note that the integral in (\ref{eq:tortoise}) cannot be done analytically for a generic function~$F(r)$, but this is not a crucial problem for the numerical computation of QNFs.}
	\begin{align}\label{eq:tortoise}
		r_* \equiv \int \sqrt{\frac{\tilde{a}_1}{a_2}}\,dr\;, 
	\end{align}
where substituting the expressions of $\tilde{a}_1$ and $a_2$ yields
    \begin{align}\label{eq:F}
    F(r)\equiv \frac{dr}{dr_*}
    =\sqrt{\frac{a_2}{\tilde{a}_1}}
    =\sqrt{\frac{B}{A}}\frac{A+\alpha_T}{\sqrt{1+\alpha_T}}\;.
    \end{align}
Also, we redefine the master variable as
	\begin{align}\label{eq:master_var}
		\Psi = (\tilde{a}_1 a_2)^{1/4}\,\chi \;.
	\end{align}
Thus, the generalized RW equation is given by
	\begin{align}\label{eq:RG}
		\frac{\partial^2 \Psi}{\partial r_*^2}  - \frac{\partial^2 \Psi}{\partial \tilde{t}^2} -  V_{\rm eff}(r) \Psi = 0 \;, 
	\end{align}
where the effective potential is defined by
	\begin{align}\label{eq:RW_potential}
		V_{\rm eff}(r) &\equiv \frac{a_4}{\tilde{a}_1} + \frac{1}{2\sqrt{\tilde{a}_1 a_2}}\frac{d^2 \sqrt{\tilde{a}_1 a_2}}{dr_*^2} - \frac{1}{4\tilde{a}_1 a_2}\bigg(\frac{d \sqrt{\tilde{a}_1 a_2}}{d r_*}\bigg)^2 \nonumber \\
		&=\sqrt{1+\alpha_T}\,F\left\{\sqrt{\frac{A}{B}}\frac{\ell(\ell+1)-2}{r^2}+\frac{r}{(1+\alpha_T)^{3/4}}\left[ F\left(\frac{(1+\alpha_T)^{1/4}}{r}\right)'\,\right]'\right\}\;.
	\end{align}
Here, we recall that a prime denotes the derivative with respect to $r$ and that we have imposed the condition~$\alpha+M_3^2=0$. Under this condition, we have $\alpha_T(r)=\alpha/(M_\star^2-\alpha)$, where $\alpha(r)$ is determined by the fourth equation in \eqref{EOM_BG} once $A(r)$ and $B(r)$ are fixed. In the case of $A(r)=B(r)$, one can use the formula~\eqref{eq:alpha2Mstar} to express $\alpha_T(r)$ as
    \begin{align}
    \alpha_T=-1-r(2-2A+rA')\frac{M_\star^2}{3\lambda}\;,
    \end{align}
with $\lambda$ being a constant of mass dimension one. Note that we have defined $r_*$ and $\Psi$, so that $\Psi$ satisfies the wave equation with unit propagation speed. Furthermore, the function~$F(r)$ defines the position of the odd-mode horizon~$r=r_g$ by $F(r_g)=0$. This, in particular, implies that the effective potential vanishes at the odd-mode horizon: $V_{\rm eff}(r_g)=0$.

Let us now briefly discuss the application of the generalized RW equation~\eqref{eq:RG} to the analysis of odd-parity gravitational waves in the time domain. Obviously, \eqref{eq:RG} is a hyperbolic equation and one can consistently set an initial condition on any hypersurface that is spacelike w.r.t.~the effective metric~$-d\tilde{t}^2+dr_*^2$. Therefore, as far as the system without the matter source is concerned, one can easily perform the time-domain analysis based on \eqref{eq:RG}. (See, e.g.,~\cite{Nakashi:2022wdg,DHOST_stealthBH_odd_prep} for related discussions in the context of the so-called Vishveshwara’s classical scattering experiment~\cite{Vishveshwara:1970zz}, i.e., the time-domain analysis with an initial Gaussian wave packet.)

In the presence of the matter source, there is a subtle point that needs to be taken into account in the time-domain analysis: A portion of a hypersurface that is spacelike w.r.t.~the effective metric can be timelike w.r.t.~the original metric to which matter fields minimally couple. Here, as a simple example, let us consider $\tilde{t}=const.$~hypersurfaces, which are spacelike w.r.t.~the effective metric, and suppose that matter minimally couples to the Einstein-frame metric~$g_{\mu\nu}$. On rewriting the master equation in the form~\eqref{eq:RG}, we had to make a redefinition of the time coordinate~$t\to\tilde{t}$ [see Eq.~\eqref{eq:tildet}]. This implies that a $\tilde{t}=const.$~hypersurface is not necessarily spacelike w.r.t.~the matter frame metric. Indeed, the norm of the vector~$\partial_\mu\tilde{t}$,
    \begin{align}\label{eq:norm_tildet}
    \bar{g}^{\mu\nu}\partial_\mu\tilde{t}\,\partial_\nu\tilde{t}
    =-\frac{A+\alpha_T(2+\alpha_T)}{(A+\alpha_T)^2}\;,
    \end{align}
can vanish at some positive $r$, which defines a critical radius where the character of a $\tilde{t}=const.$~hypersurface changes. Therefore, in the presence of the matter source, one needs to carefully choose the initial hypersurface so that it is spacelike not only w.r.t.~the effective metric but also w.r.t.~the matter-frame metric. Obviously, the $\tau=const.$~hypersurfaces in the unitary gauge, i.e., the hypersurfaces on which the scalar field in the gravity sector is constant, are candidates for consistent initial hypersurfaces since the scalar field is expected to play the role of a clock and thus defines the causality in the sense of the past and the future. Once the initial condition is prepared on a consistent initial hypersurface, simulations with the matter source can be performed in a straightforward way. Then, the way how matter excites QNMs is expected to be different from GR. It is certainly of great interest to perform the time-domain analysis in the presence of the matter source.

\section{Quasinormal mode spectrum}\label{sec:QNM}

In the previous Section, we have derived the generalized RW equation which describes the dynamics of linear perturbations in the odd-parity sector. In this Section, we are going to determine the spectrum of QNMs. In particular, we will solve Eq.~(\ref{eq:RG}) with appropriate boundary conditions at $r = r_g$ and at $r \rightarrow \infty$. 

Throughout this Section, we use the following ansatz for $\Psi$: $\Psi(\tilde{t}, r_*) = Q(r_*) e^{-i\omega \tilde{t}}$ with $\omega$ being the frequency of the mode function~$Q(r_*)$. Plugging such an ansatz into Eq.~(\ref{eq:RG}), we obtain
\begin{align}\label{eq:Q_master}
\frac{d^2}{dr_*^2}Q(r_*) + (\omega^2 - V_{\rm eff})Q(r_*) = 0 \;.
\end{align}
This is the main equation we solve to determine the spectrum of QNMs. The QNMs are defined by solutions to the above equation such that $Q(r_*)\sim e^{-i\omega r_*}$ (purely ingoing) in the vicinity of the odd-mode horizon~$r=r_g$ (or equivalently $r_*\to-\infty$) and $Q(r_*)\sim e^{i\omega r_*}$ (purely outgoing) as $r\to\infty$ (or $r_*\to\infty$). Actually, the equation above resembles the one-dimensional time-independent Schr{\"o}dinger equation with the potential~$V_{\rm eff}(r_*)$. In this sense, our problem simply becomes a scattering problem of a wave with the effective potential provided that the wave is purely ingoing as $r_* \rightarrow -\infty$ and purely outgoing as $r_* \rightarrow \infty$. Actually, it is interesting to mention that the spectrum for the scattering problem above is related to the one for the bound state problem via the analytic continuation of $r_*$: $r_* \rightarrow -i r_*$, resulting in the relation~$\omega^2 = -E$, where $E$ is the bound state energy. See \cite{Hatsuda:2021gtn} and references therein for a detailed discussion of this method.\footnote{In fact, this approach is particularly useful for computing QNFs of a system where the spectrum of the bound states can be analytically obtained. Recently, in \cite{Volkel:2022ewm}, it was shown that the bound state method can also be applied to potentials where the spectrum can be computed only numerically.}

In what follows, we will demonstrate the computation of QNFs in two cases: One is the stealth Schwarzschild solution, and the other is the Hayward solution.

\subsection{Stealth Schwarzschild solution}
As a warm-up, in this Subsection, we consider the so-called stealth solutions where the parameter~$\alpha_T$ is a constant (or equivalently $M_3^2=const.$). As pointed out before, this parameter~$\alpha_T$ modifies the speed of GWs to be different from that of light. Therefore, one needs to take into account the LIGO/Virgo bound~\cite{TheLIGOScientific:2017qsa,GBM:2017lvd,Monitor:2017mdv} when the length scale of GWs enters the observational regime, which requires essentially $|\alpha_T|\lesssim 10^{-15}$. Having said that, for illustrative purposes, we will also consider $\alpha_T$ of ${\cal O}(10^{-1})$ to demonstrate how the QNM spectrum gets modified.

A stealth black hole solution was first found in the context of k-essence and ghost condensation in \cite{Mukohyama:2005rw}. Then, the general construction of stealth solutions was developed in \cite{Motohashi:2018wdq,Takahashi:2020hso} and perturbations around such solutions have been studied, e.g., in \cite{Babichev:2018uiw,Takahashi:2019oxz,deRham:2019gha,Motohashi:2019ymr,Khoury:2020aya,Tomikawa:2021pca,Takahashi:2021bml}. In the case of DHOST theories, perturbations about the stealth solutions are known to be strongly coupled, but this problem can be cured by a small detuning (i.e., \emph{scordatura}) of the degeneracy condition~\cite{Motohashi:2019ymr} if and only if the scalar field profile is timelike. Interestingly, the scordatura effect is built-in in ghost condensation~\cite{Arkani-Hamed:2003pdi} and U-DHOST theories~\cite{DeFelice:2022xvq}, and approximately stealth black hole solutions in the presence of the scordatura term were studied in \cite{Mukohyama:2005rw,Cheng:2006us,DeFelice:2022qaz}. Since our EFT encompasses the U-DHOST class (as well as ghost condensation), we expect that a similar thing should happen in our EFT: The scordatura effect would make the perturbations weakly coupled, while the stealth black hole background would be subjected to a small correction controlled by the scordatura term. Having said that, in the odd-mode analysis, the scordatura effect would be negligible and we employ exactly stealth configurations as our background for simplicity.

\begin{figure}[t]
    \centering
    \includegraphics[width=0.6\linewidth]{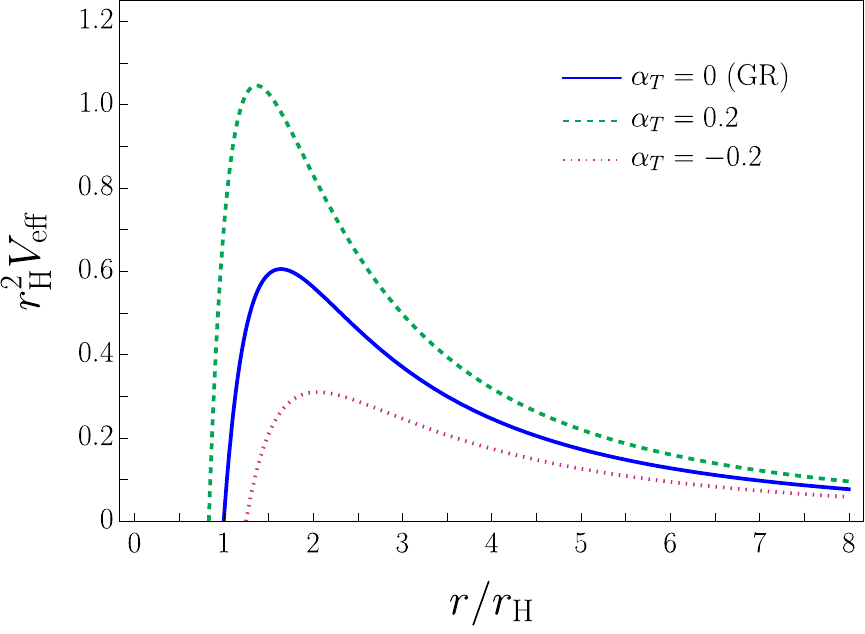}
    \caption{~The effective potential~(\ref{eq:RW_potential_const_alpha3}) as a function of $r/r_{\rm H}$ in the 
    stealth-Schwarzschild case for $\ell = 2$. The blue solid line corresponds to $\alpha_T = 0$ (the standard RW potential in GR), while the green dashed and the pink dotted lines correspond to  $\alpha_T = 0.2$, and $-0.2$ respectively. The potentials drop to zero when $r = r_g$ with  
    $r_g\equiv r_{\rm H}/(1+\alpha_T)$ and fall off to zero as $r \rightarrow \infty$.} 
    \label{fig:RW_potential}
\end{figure}

In what follows, we consider the case where the metric is given by the Schwarzschild one, for which the metric is described by
\begin{align}\label{eq:A_B_const_alpha3}
	A(r) = B(r) = 1 - \frac{r_{\rm H}}{r} \;,
	\end{align}
with $r_{\rm H}$ denoting the photon horizon. Note that Eq.~\eqref{eq:alpha2Mstar}, derived from one of the components of the background equations, is automatically satisfied in this case with $\alpha-M_\star^2=const$. The effective potential~(\ref{eq:RW_potential}) for generic multipole index~$\ell$ takes the following form: 
\begin{align}\label{eq:RW_potential_const_alpha3}
V_{\rm eff}(r) = 
(1+\alpha_T)f(r)\left[\frac{\ell(\ell+1)}{r^2} - \frac{3r_g}{r^3} \right] \;,
\end{align}
where $f(r) \equiv 1 - r_g/r$, with $r_g\equiv r_{\rm H}/(1+\alpha_T)$ being the odd-mode graviton horizon. As expected, when $\alpha_T = 0$, the potential above reduces to the standard RW potential in GR. In Figure~\ref{fig:RW_potential}, we plot the potential~(\ref{eq:RW_potential_const_alpha3}) as a function of $r/r_{\rm H}$ for $\alpha_T = 0$, $0.2$, and $-0.2$. In the plot, the blue solid curve corresponds to the GR limit ($\alpha_T = 0$), whereas the green dashed and the pink dotted curves represent the modification of gravity with $\alpha_T = 0.2$ and $-0.2$, respectively. In comparison with the GR potential (blue solid), the potential gets higher and narrower for $\alpha_T>0$, while it gets lower and wider for $\alpha_T<0$. We also see that the potential drops to zero at $r=r_g$.

It is interesting to point out that for positive $\alpha_T$ the graviton horizon is smaller than the photon horizon, which might lead to some interesting phenomenology due to the fact that the graviton modes become superluminal, if $\alpha_T$ is promoted to a function of $r$ and vanishes at $r\to\infty$ to satisfy the LIGO/Virgo bound. We leave the study of such a case with $r$-dependent $\alpha_T$ to the next subsection and, for simplicity, restrict our consideration in this subsection to the case of constant $\alpha_T$. For completeness, in this Subsection, we will compute the QNFs for both positive and negative $\alpha_T$.\footnote{Note that, in the present case, the critical radius within which a constant-$\tilde{t}$ hypersurface becomes timelike can be calculated with \eqref{eq:norm_tildet} as $r=r_{\rm H}/(1+\alpha_T)^2$, which is larger than $r_g=r_{\rm H}/(1+\alpha_T)$ for $(-1<)\,\alpha_T<0$.} Moreover, in the constant-$\alpha_T$ case, the integral in \eqref{eq:tortoise} can be done analytically, giving rise to $r_*(r)$ as 
	\begin{align}\label{eq:tortoise_gen}
	r_*(r) = (1+\alpha_T)^{-1/2}\left[r + r_g \log\left|\frac{r}{r_g} - 1\right|\,\right] \;.
	\end{align}
Note that this tortoise coordinate~$r_*$ goes to $-\infty$ as $r \rightarrow r_g$, while it goes to $+\infty$ as $r \rightarrow \infty$. 

Let us now study QNFs for the stealth Schwarzschild solution. Interestingly, in the present case, our master equation~\eqref{eq:Q_master} can be recast into the standard RW equation in GR by the following rescalings:
    \begin{align}
    \begin{split}
    \tilde{\omega}&\equiv (1+\alpha_T)^{-1/2}\,\omega\;, \\
    \tilde{r}_*&\equiv \sqrt{1+\alpha_T}\,r_*
    =r + r_g \log\left|\frac{r}{r_g} - 1\right|\;, \\
    \tilde{V}_{\rm eff}&\equiv \frac{V_{\rm eff}}{1+\alpha_T}
    =f(r) \left[\frac{\ell(\ell+1)}{r^2} - \frac{3r_g}{r^3} \right]\;.
    \end{split}
    \end{align}
It should be noted that $\tilde{r}_*$ and $\tilde{V}_{\rm eff}$ are nothing but the standard tortoise coordinate and the RW potential in GR with the horizon radius~$r_g$ (rather than $r_{\rm H}$), respectively. Written explicitly, we have 
\begin{align}\label{eq:tilde_const}
\frac{d^2}{d\tilde{r}_*^2}Q + (\tilde{\omega}^2 - \tilde{V}_{\rm eff})Q = 0 \;.
\end{align}
This implies that there is a simple relation between the QNM spectrum for the stealth Schwarzschild solution in our EFT and the one for the Schwarzschild solution in GR:
    \begin{align}
    r_g \tilde{\omega} = r_{\rm H} \omega_{\rm GR}\;.
    \end{align}
Hence, the QNFs normalized by $r_{\rm H}$ can be written as
\begin{align}\label{eq:QNM_const_3_rescaled}
r_{\rm H} \omega = r_{\rm H} \omega_{\rm GR} 
(1 + \alpha_T)^{3/2}\;,
\end{align}
where we have used $r_g=r_{\rm H}/(1+\alpha_T)$. In GR, QNFs for the odd-parity modes (normalized by $r_{\rm H}$) are known in the literature (see, e.g., \cite{Chandrasekhar:1975zza,Schutz:1985km} for early works). Using the GR values and applying the formula~(\ref{eq:QNM_const_3_rescaled}), we obtain QNFs of the stealth-Schwarzschild case with constant $\alpha_T$. Notice that the expression~(\ref{eq:QNM_const_3_rescaled}) holds not only for the fundamental QNM but also for any overtones of the QNMs. 

As a check, we perform the so-called direct integration method (see, e.g., \cite{Molina:2010fb}) in {\texttt{Mathematica}} to determine the fundamental QNF for $\ell=2$, which we denote by $\omega_0$. In order to use such a method, as mentioned earlier, one needs to impose the following boundary conditions: 
\begin{align}
Q(r_*) \propto \left\{
\begin{matrix}
e^{-i \omega r_*} \;, & r_* \rightarrow -\infty & (r \rightarrow r_g) \\
e^{i \omega r_*} \;, & r_* \rightarrow \infty & (r \rightarrow \infty)
\end{matrix}
\right. \;.
\end{align}
Notice that it is important to impose the boundary condition at the graviton horizon~$r_g$, not at the photon horizon. After that, one numerically solves the equation of motion starting from the horizon~$r_g$ and from infinity, to some arbitrary matching point~$r=r_{\rm m}$. This yields a piecewise solution, say, $Q=Q_-$ for $r<r_{\rm m}$ and $Q=Q_+$ for $r>r_{\rm m}$. Then, in order to obtain the QNF, one requires that $Q$ and its first derivative are continuous at $r=r_{\rm m}$. Actually, this continuity condition can be satisfied only for some particular values of $\omega$ corresponding to QNFs, among which the one with the smallest (absolute value of the) imaginary part gives the fundamental QNF~$\omega_0$.\footnote{In practice, this can be done by the root-finding algorithm implemented in \texttt{Mathematica}, where one has to choose some trial value for $\omega$. In the present case, we can give the trial value by the analytic formula~\eqref{eq:QNM_const_3_rescaled} to confirm that the direct integration method yields a consistent result. If, on the other hand, an analytic formula of QNFs cannot be obtained, one can use, for instance, the WKB approximation to determine a naive estimate of QNFs and use it as the trial value for the direct integration method. Note that the WKB method can be accurately applied to the case in which the potential has only a single peak.} We confirm that our numerical results are robust against the changes of the matching location~$r_{\rm m}$ and the step size one uses to solve the master equation~\eqref{eq:Q_master} both from $r_g$ and from infinity.

\begin{figure}[t!]
	\begin{subfigure}{0.48\textwidth}
		\includegraphics[width=\linewidth]{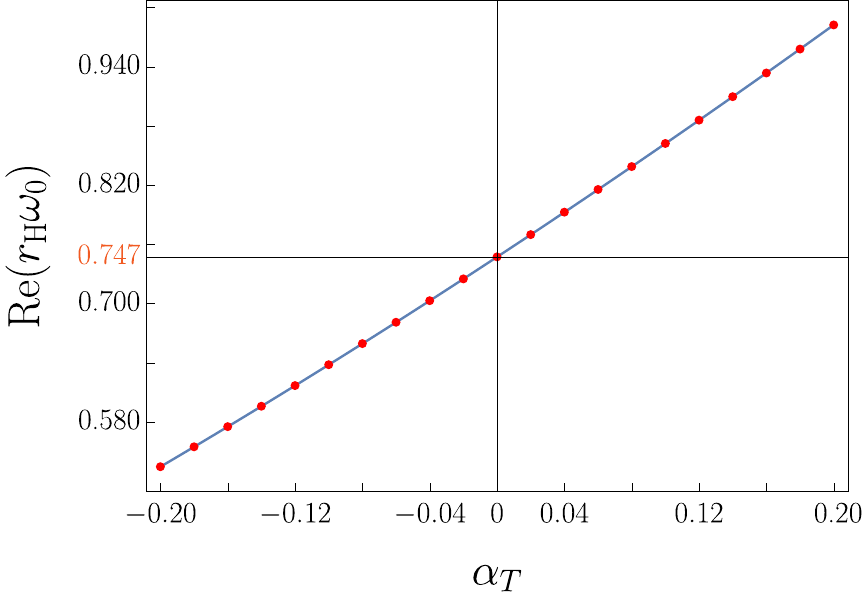}
	\end{subfigure}
	\hspace*{\fill} 
	\begin{subfigure}{0.48\textwidth}
		\includegraphics[width=\linewidth]{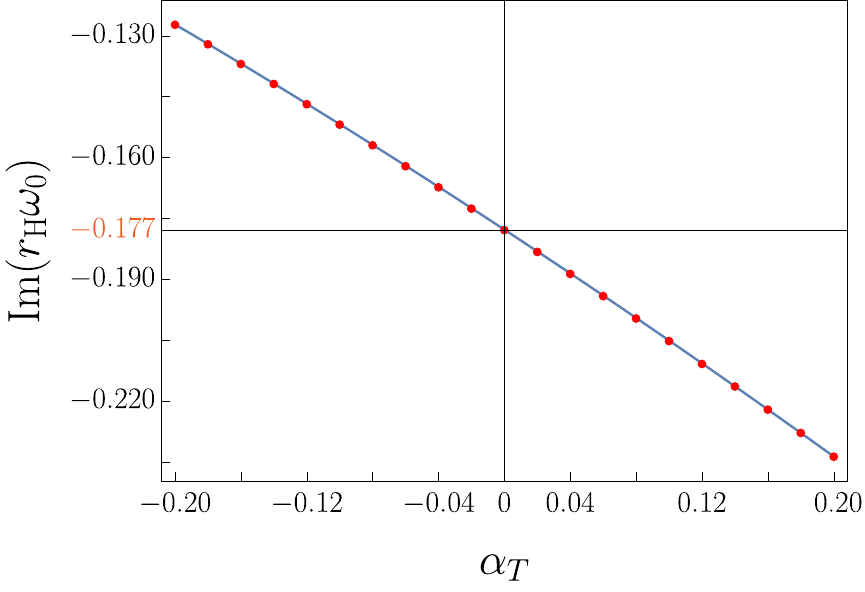}
	\end{subfigure}
	\hspace*{\fill} 
	\caption{~The comparison of the fundamental QNFs for both real (left panel) and imaginary (right panel) parts, using the analytic formula (blue curves) given by Eq.~(\ref{eq:QNM_const_3_rescaled}) and the numerical values (red data points) obtained by using the direct integration method for $\alpha_T \in [-0.2,0.2]$. The red numbers on the vertical axes in both plots indicate the GR values.} 
	\label{fig:Comparison_anal_Direct}
\end{figure}

The comparison of the fundamental QNFs obtained by the analytic formula~(\ref{eq:QNM_const_3_rescaled}) and by the direct integration method is shown in Figure~\ref{fig:Comparison_anal_Direct}. Notice that here we focus on the fundamental modes (the least-damped modes) which are expected to dominate the late-time ringdown phase, although the importance of higher overtones is still a debating subject, see, e.g., \cite{Nee:2023osy,Giesler:2019uxc}. Furthermore, we also use the WKB approximation at sixth-order~\cite{Iyer:1986np,Konoplya:2003ii} to compute the fundamental QNFs. The numerical values of the fundamental QNFs using the methods mentioned above are presented in Table~\ref{Table:QNMs_const_alpha3} for several values of $\alpha_T$. We see that the results from the analytic formula and the direct integration method perfectly coincide at least up to sixth decimal place, while the sixth-order WKB approximation also shows a good agreement with the other two methods. Therefore, we confirm that the QNFs obtained from the formula~(\ref{eq:QNM_const_3_rescaled}) is very robust and can be used for further applications. More importantly, this agreement validates the numerical codes used for the computation. 

\begin{table}
  \centering 
  \begin{threeparttable}
    \begin{tabular}{cccc}
    $\alpha_T$  & Analytic formula & Direct integration & Sixth-order WKB  \\
    \midrule\midrule
   $-0.12$  &  $0.616942 - 0.146879 i$ & $0.616942 - 0.146879 i$ &  $0.616855 - 0.146761 i$  \\ 
      \cmidrule(l  r ){1-4}
    $-0.08$  &  $0.659480 - 0.157007 i$ & $0.659480 - 0.157007 i$ &  $0.659388 - 0.156881 i$  \\ 
     \cmidrule(l  r ){1-4}
   $-0.04$  &  $0.702954 - 0.167357 i$ & $0.702954 - 0.167357 i$ &  $0.702856 - 0.167222 i$  \\ 
    \cmidrule(l  r ){1-4}
  $0$ (GR) &   ($0.747343 - 0.177925 i$) &  $0.747343 - 0.177925 i$   &  $0.747239 - 0.177782 i$   \\
    \cmidrule(l  r ){1-4}
   $0.04$  &  $0.792629 - 0.188706 i$ & $0.792629 - 0.188706 i$ &  $0.792518 - 0.188555 i$  \\ 
    \cmidrule(l r ){1-4}
     $0.08$ & $0.838795 - 0.199697 i$ & $0.838795 - 0.199697 i$  & $0.838678 - 0.199537 i$ \\ 
    \cmidrule(l r ){1-4}
    $0.12$ & $0.885824 - 0.210893 i$ & $0.885824 - 0.210893 i$   & $0.885699 - 0.210724 i$   \\
     \midrule\midrule
    \end{tabular}
  \end{threeparttable}
  \caption{~The fundamental QNFs of the modes with $\ell = 2$ for several values of $\alpha_T$, which is a constant for the stealth Schwarzschild solution. We use three methods to compute the frequencies: the analytic formula~(\ref{eq:QNM_const_3_rescaled}), the direct integration method, and the sixth-order WKB approximation. On applying the analytic formula, we use the GR value known in the literature as a reference value, which is indicated by the parenthesis.} \label{Table:QNMs_const_alpha3}
  \end{table}
 
Although the QNFs we obtained in this Subsection differ from those of GR (corresponding to $\alpha_T = 0$), the analytic formula~\eqref{eq:QNM_const_3_rescaled} implies that the real and the imaginary parts of the QNFs are rescaled by the same factor~$(1+\alpha_T)^{3/2}$, and hence the ratio between them is indistinguishable from the GR value.
Note that such a rescaling of QNFs, which holds for any multipoles and overtones, can be absorbed into a redefinition of the black hole mass.
Therefore, in the case of the stealth Schwarzschild solution, one cannot determine the value of $\alpha_T$ with observations of QNFs unless the black hole mass is known from other observations.
This situation could change for rotating black holes (e.g., the stealth Kerr solution~\cite{Charmousis:2019vnf,Takahashi:2020hso}), which is however beyond the scope of the present paper.
Another possible way to resolve the degeneracy is to study the QNFs in the even-parity sector.
Since the perturbation of the scalar field is absent in the odd-parity mode and only contributes to the even-parity mode, the even-parity QNFs are expected to differ from those of GR.
We leave this analysis to future work.

In the next Subsection, we are going to investigate the case where $\alpha_T$ is a non-trivial function of $r$ that asymptotically goes to zero as $r \rightarrow \infty$, so that it is consistent with LIGO/Virgo bound. In fact, this might be phenomenologically interesting in the sense that the modification of gravity only shows up in the vicinity of the black hole and could leave some observational imprints on QNFs.

\subsection{Hayward solution}\label{sec:Hayward}

Here we study, as an illustrative example of non-stealth metric profiles, the Hayward metric~\cite{Hayward:2005gi}, for which
    \begin{align}
    A(r) = B(r) =1-\frac{\mu r^2}{r^3+\sigma^3}\;,
    \end{align}
with $\mu\,(>0)$ and $\sigma$ being parameters of length dimension. For $\sigma>0$, the metric approaches the Schwarzschild metric as $r\to \infty$ and the de Sitter metric (with the effective cosmological constant given by $3\mu/\sigma^3$) as $r\to 0$, so that there is no curvature singularity. For $\sigma<0$, the Hayward metric does not describe a regular black hole as there exists a curvature singularity at $r=-\sigma\,(>0)$, but one could regard $\sigma$ as just a phenomenological parameter that controls the deviation from the Schwarzschild metric, and hence we keep this possibility open. When the regularization scale~$\sigma$ vanishes, we recover the Schwarzschild metric. Usually, in order to realize such a geometry as a solution in GR, one has to introduce some non-standard matter field, e.g., nonlinear electrodynamics~\cite{IlichKruglov:2021pdw}. In our EFT, one can realize this geometry as an exact background solution by choosing the coefficients of the tadpole terms appropriately (see Subsection~\ref{sec:background}). In particular, the condition~\eqref{eq:alpha2Mstar} fixes the function~$\alpha(r)$ in the form
    \begin{align}\label{eq:alpha2Mstar_Hayward}
    \alpha=M_\star^2+\frac{\lambda(r^3+\sigma^3)^2}{\mu r^6}\;,
    \end{align}
with $\lambda$ being a constant.

The position of the horizon(s) for photons is given by the solution to the following cubic algebraic equation:
    \begin{align}
    r^3-\mu r^2+\sigma^3=0\;,
    \label{eq:r_p-Hayward}
    \end{align}
which has two solutions~$r_\pm$ with $r_+>r_-$ (outer and inner horizons, respectively) in the region~$r>0$ if $\mu$ and $\sigma$ satisfy the following condition:
    \begin{align}
    0<\frac{\sigma^3}{\mu^3}<\frac{4}{27}\;.
    \end{align}
As a side note, one can verify that $r_+<\mu$. The two horizons are degenerate for $\sigma^3/\mu^3=4/27$. Also, in the limit~$\sigma\to 0$, we have $r_+\to \mu$ and $r_-\to 0$.\footnote{When $\mu>0$ and $\sigma<0$, Eq.~\eqref{eq:r_p-Hayward} has only one positive solution.} 

\begin{figure}[t]
    \centering
    \includegraphics[width=0.6\linewidth]{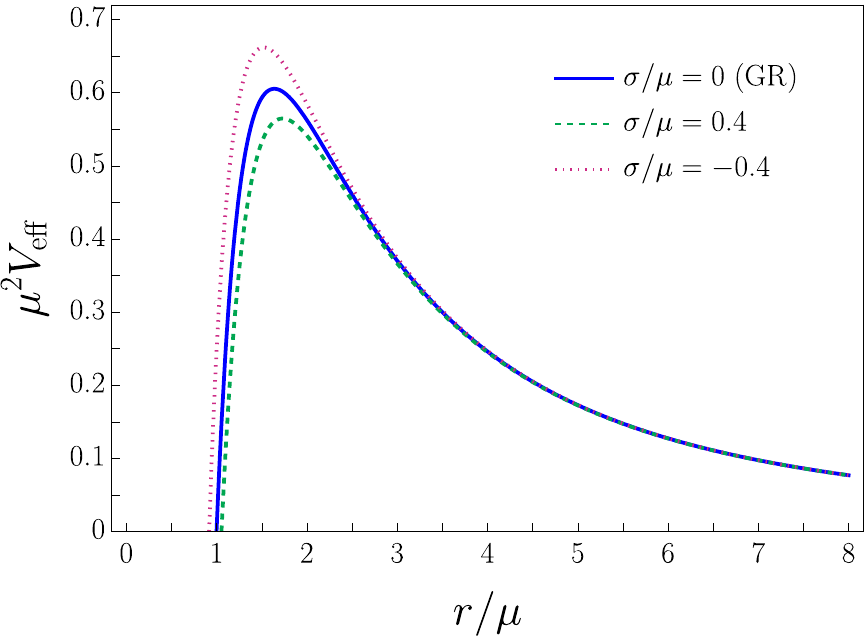}
    \caption{~The effective potential~\eqref{eq:Hayward_potential} associated with the Hayward metric for $\ell=2$. The blue solid line corresponds to $\sigma/\mu = 0$ (the RW potential in GR), whereas the green dashed and the pink dotted lines correspond to $\sigma/\mu = 0.4$, and $-0.4$ respectively. The potentials, as usual, drop to zero when $r = r_g$ with $r_g/\mu$ satisfying the condition (\ref{eq:r_g-Hayward}), and fall off to zero when $r \rightarrow \infty$.}
    \label{fig:Hayward_potential}
\end{figure}

Let us now consider odd-parity perturbations about the Hayward metric based on our EFT framework. In order to satisfy the LIGO/Virgo bound, we would like to have $\alpha_T=c_T^2-1\to 0$ (or equivalently $M_3^2=-\alpha\to 0$) at spatial infinity. This requirement poses the following relation between $\lambda$ and $\mu$:
    \begin{align}
    \lambda+M_\star^2\mu=0\;,
    \end{align}
which yields
    \begin{align}
    \alpha_T(r)\equiv -\frac{M_3^2}{M_\star^2+M_3^2}
    =-\frac{\sigma^3(2r^3+\sigma^3)}{(r^3+\sigma^3)^2}\;.
    \end{align}
Here, we have also used Eq.~\eqref{eq:alpha2Mstar_Hayward}. The function~$F\equiv dr/dr_*$ now reads [see Eq.~\eqref{eq:F}] 
    \begin{align}\label{eq:drdrs-Hayward}
    F(r)=\frac{r^4 - \mu (r^3 + \sigma^3)}{r (r^3 + \sigma^3)}\;.
    \end{align}
The position of the odd-mode horizon~$r_g\,(>0)$ is given by $F(r_g)=0$, or equivalently,
    \begin{align}
    r_g^4-\mu(r_g^3+\sigma^3)=0\;,
    \label{eq:r_g-Hayward}
    \end{align}
which has a single positive solution so long as $\mu$ and $\sigma$ are positive.\footnote{When $\mu>0$ and $\sigma<0$, Eq.~\eqref{eq:r_g-Hayward} has two positive solutions for $-27/256<\sigma^3/\mu^3\,(<0)$, while no positive solution exists for $\sigma^3/\mu^3<-27/256$. Note also that, for $\sigma>0$, a constant-$\tilde{t}$ hypersurface can become timelike in the vicinity of $r=r_g$, and the critical radius is given by the unique positive solution to $r^{10}-\mu(r^3+\sigma^3)^3=0$.} This relation determines the value of $r_g/\mu$ once we fix $\sigma/\mu$. As a side note, one can verify that $r_g>\mu\,(>r_+)$. In practice, it is useful to regard $r_g$ and $\sigma$ as independent parameters and fix $\mu$ so that
    \begin{align}
    \mu=\frac{r_g^4}{r_g^3+\sigma^3}\;.
    \end{align}
Then, from Eq.~\eqref{eq:RW_potential}, one obtains the effective potential as
    \begin{align}
    V_{\rm eff}(r)
    =\left[1-\frac{\mu(r^3+\sigma^3)}{r^4}\right]
    \left\{\frac{\ell(\ell+1)r^4}{(r^3+\sigma^3)^2}-\frac{3\left[4\mu r^9+2\sigma^3r^6(8r-\mu)+\sigma^6r^3(r-7\mu)-\mu\sigma^9\right]}{4(r^3+\sigma^3)^4}\right\}\;.
    \label{eq:Hayward_potential}
    \end{align}
As mentioned earlier, the effective potential vanishes at the odd-mode horizon ($r=r_g$) and at spatial infinity. Note also that this effective potential reduces to the standard RW potential for $\sigma/\mu = 0$. We plot the above $V_{\rm eff}(r)$ in Figure~\ref{fig:Hayward_potential}. In the plot, the blue solid curve corresponds to $\sigma/\mu = 0$ (the standard RW potential in GR), while the green dashed and the pink dotted curves correspond to $\sigma/\mu = 0.4$ and $-0.4$, respectively. In comparison with the RW potential in GR (blue solid), the potential gets higher and narrower for $\sigma/\mu < 0$, while it gets lower and wider for $\sigma/\mu > 0$. 

\begin{figure}[t]
    \centering
    \includegraphics[width=0.95\linewidth]{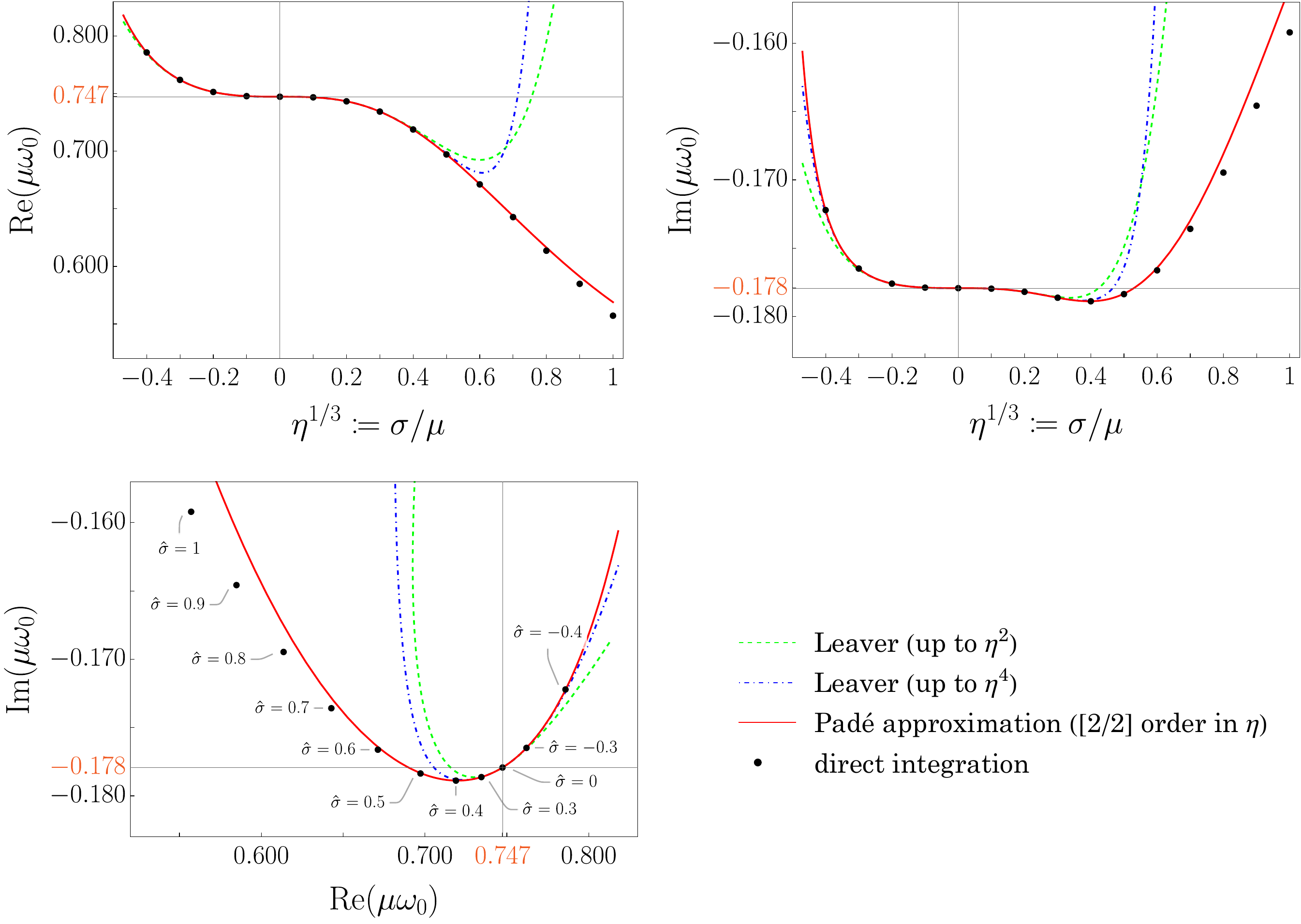}
    \caption{~The $\ell=2$ fundamental QNF~$\omega_0$ (normalized by $\mu$) for the Hayward solution. The black dots represent the result of direct integration method. The green dashed and the blue dash-dotted curves correspond to the Leaver series expansion \eqref{QNM_Hayward_Leaver}, up to order $\mathcal{O}(\eta^2)$ and $\mathcal{O}(\eta^4)$ respectively. The red curve represents the Pad{\'e} approximation \eqref{QNM_Hayward_Pade} of order $[2/2]$ in $\eta$. The top panels show the real and imaginary parts of $\omega_0$ as a function of $\hat{\sigma}\equiv \sigma/\mu\,(=\eta^{1/3})$. The bottom left panel shows the trajectory of $\omega_0$ in the complex $\mu\omega_0$ plane.}
    \label{fig:QNM_Hayward}
\end{figure}

Let us now proceed with the computation of QNFs. As mentioned in the previous Subsection, we only focus on the fundamental modes. By employing the direct integration method, we obtain the fundamental QNFs, $\omega_0$, for $\ell=2$ as a function of $\hat{\sigma}\equiv \sigma/\mu$. The numerical values of fundamental QNFs for different values of $\hat{\sigma}$ are displayed in Table~\ref{Table:QNMs_Hayward} and are shown as black dots in Figure~\ref{fig:QNM_Hayward}. As mentioned earlier, in order to apply the direct integration method, one has to choose a trial value for $\omega$. One can use the GR value as the trial value if $|\hat{\sigma}|\ll 1$, otherwise this may not be the case. If, on the other hand, one wants to compute the QNF for positive $\hat{\sigma}$ of ${\cal O}(1)$ or larger, then the strategy is as follows. (A similar strategy applies to negative $\hat{\sigma}$.) First, one starts with a small positive $\hat{\sigma}$, say $\hat{\sigma}=0.1$, where a QNF can be found by using the GR value of $\omega_0$ as the trial value. Then, one proceeds to a slightly larger $\hat{\sigma}$, where one can use the QNF obtained for the previous value of $\hat{\sigma}$ as the trial value. By repeating this procedure, one can obtain a QNF for $\hat{\sigma}={\cal O}(1)$ or larger, keeping track of the (would-be) fundamental QNF.\footnote{Strictly speaking, for large $\hat{\sigma}$, there is no guarantee that the so-obtained QNF corresponds to the actual fundamental QNF. In order to confirm this point, one has to study the full QNM spectrum by some other method.} We also confirm that our numerical result is stable under varying the values of the matching point~$r_{\rm m}$ and the step size used to solve the differential equation both from $r_g$ and from infinity.  

As a cross-check, we apply the explicit sixth-order WKB formula to compute the fundamental QNFs. In Table~\ref{Table:QNMs_Hayward}, we show the frequencies obtained from the WKB approximation. This indicates that the WKB method accurately reproduces the result of the direct integration method up to $\hat{\sigma} \sim \mathcal{O}(1)$. Notice that even for $\hat{\sigma} \gtrsim 0.6$ the results of the two methods coincide at least up to second decimal place.

Furthermore, we employ Leaver's continued fraction method~\cite{Leaver:1985ax} with Nollert's improvement~\cite{Nollert:1993zz,Zhidenko:2006rs}, which allows us to compute the QNF analytically. (See Appendix~\ref{sec:Leaver} for a more detailed explanation.) Assuming that the QNF for the Hayward solution is expressed in the form of
    \begin{align}
    \omega_0(\eta)=\omega_{0,{\rm GR}}^{(\ell=2)}+\omega_{0}^{(1)}\eta+\omega_{0}^{(2)}\eta^2+\omega_{0}^{(3)}\eta^3+\omega_{0}^{(4)}\eta^4+\cdots\;,
    \end{align}
with $\eta\equiv \hat{\sigma}^3$, one can determine the expansion coefficients order by order in $\eta$. Up to ${\cal O}(\eta^4)$, the QNF can be written as
    \begin{align}
    \mu \omega_0&=(0.747343 - 0.177925 i) - (0.507532 + 0.0372539 i) \eta + (1.17402 + 
    0.491105 i) \eta^2 \nonumber \\
    &\quad - (3.61571 + 
    3.09972 i) \eta^3 + (11.6504 + 18.6304 i) \eta^4 + {\cal O}(\eta^5)\;, \label{QNM_Hayward_Leaver}
    \end{align}
where we have used the GR value~$\omega_{0,{\rm GR}}^{(\ell=2)}$ known in the literature as an input. Note that $\mu$ amounts to twice the ADM mass (times the gravitational constant, precisely). In Figure~\ref{fig:QNM_Hayward}, the green dashed and the blue dash-dotted curves, which correspond to the expansion of $\mathcal{O}(\eta^2)$ and $\mathcal{O}(\eta^4)$ respectively, precisely reproduce the result of direct integration method up to $\hat{\sigma}\simeq 0.4$.\footnote{Another useful semi-analytic method to compute the QNF is the so-called parametrized QNM ringdown formalism~\cite{Cardoso:2019mqo,McManus:2019ulj,Kimura:2020mrh,pQNM_prep}. We confirm that the series expansion~\eqref{QNM_Hayward_Leaver} coincides with the result of parametrized QNM formalism up to ${\cal O}(\eta^2)$.} Moreover, based on the series expansion \eqref{QNM_Hayward_Leaver}, one can obtain the Pad{\'e} approximant of $[2/2]$ order in $\eta$ as
    \begin{align}
    \mu \omega_0=\frac{(0.747343 - 
   0.177925 i) + (4.37500 + 0.291186 i) \eta + (2.85750 + 
    1.14870 i) \eta^2}{1 + (6.08372 + 
    1.88787 i) \eta + (5.39722 + 3.75020 i) \eta^2}+{\cal O}(\eta^5)\;, \label{QNM_Hayward_Pade}
    \end{align}
whose series expansion around $\eta=0$ coincides with \eqref{QNM_Hayward_Leaver}. As shown in Figure~\ref{fig:QNM_Hayward}, the Pad{\'e} approximant shows an excellent agreement with the result of direct integration method up to $\eta={\cal O}(1)$. Notice that, for $\hat{\sigma} \gtrsim 0.6$, the numerical results of both methods agree with each other at least up to second decimal digit. 

Before closing this Subsection, let us comment on the 
distinguishability of the above results from those in GR.
It is actually straightforward to deduce from the bottom left panel of Figure~\ref{fig:QNM_Hayward} that the ratio between the real and the imaginary parts of the fundamental QNF depends on the parameter~$\hat{\sigma}$, which is in sharp contrast to the case of the stealth Schwarzschild solution.
This suggests that, if one considers a black hole described by the Hayward solution, then one can in principle determine the black hole mass and the value of $\hat{\sigma}$ simultaneously from the fundamental QNF.\footnote{Of course, studying overtones would provide more information on the modification of gravity in the vicinity of the black hole.}
Moreover, as mentioned earlier, the QNM spectrum in the even-parity sector would be another interesting probe for modified gravity effects. 
Finally, to make an actual connection to observations, a more sophisticated analysis such as Bayesian analysis is required. We leave all of these investigations to future work.
\begin{table}
  \centering 
  \begin{threeparttable}
    \begin{tabular}{cccc}
    $\hat{\sigma}$ & Direct integration & Leaver \& $[2/2]$-order Pad{\'e} & Sixth-order WKB\\
    \midrule\midrule
    $-0.4$ & $0.785794 - 0.172211 i$ & $0.785798 - 0.172238 i$ & $0.786026 - 0.171832 i$ \\ 
    \cmidrule(l r ){1-4}
    $-0.2$ & $0.751481 - 0.177594 i$ & $0.751481 - 0.177594 i$ & $0.751383 - 0.177462 i$ \\ 
    \cmidrule(l r ){1-4}
    $0$ (GR) & $0.747343 - 0.177925 i$ & ($0.747343 - 0.177925 i$) & $0.747239 - 0.177782 i$\\ 
    \cmidrule(l r ){1-4}
    $0.2$ & $0.743356 - 0.178193 i$ & $0.743356 - 0.178193 i$ & $0.743253 - 0.178030 i$\\ 
    \cmidrule(l r ){1-4}
    $0.4$ & $0.718887 - 0.178884 i$ & $0.718889 - 0.178881 i$ & $0.718975 - 0.178378 i$\\ 
    \cmidrule(l r ){1-4}
    $0.6$ & $0.671222 - 0.176619 i$ & $0.671440 - 0.176435 i$ & $0.672625 - 0.174153 i$\\
    \cmidrule(l r ){1-4}
    $0.8$ & $0.613627 - 0.169466 i$ & $0.616429 - 0.168101 i$ & $0.617609 - 0.161954 i$\\
    \cmidrule(l r ){1-4}
    $1$ & $0.557171 - 0.159204 i$ & $0.568938 - 0.155897 i$ & $0.561707 - 0.144351 i$\\
    \midrule\midrule
    \end{tabular}
 \end{threeparttable}
    \caption{~The fundamental QNFs of the modes with $\ell = 2$ for different values of $\hat{\sigma}\equiv \sigma/\mu$. We use the direct integration, the Leaver's method with Nollert's improvement and the WKB approximation.  For the second method, we use the GR value known in the literature as a reference value and applied the Pad{\'e} approximation up to $[2/2]$ order in $\eta\equiv \hat{\sigma}^3$. The number in the parenthesis indicates the reference value we use in the second method.}
    \label{Table:QNMs_Hayward}
  \end{table}

	
\section{Conclusions} \label{sec:conclusions}

Studying the QNFs of black holes is an essential step towards testing gravity with the ringdown gravitational wave signal. In the present paper, we have investigated the QNFs based on the EFT with a timelike scalar profile~\cite{Mukohyama:2022enj} applied to a static and spherically symmetric black hole background. In particular, we have established a pipeline to compute the QNFs by writing down the master equation in a ready-to-use form and implementing the analytical/numerical methods for the QNF computation.

In Section~\ref{sec:QNM}, we have demonstrated the computation of the QNFs with two examples of hairy black holes: One is the stealth Schwarzschild solution and the other is the Hayward metric with a non-trivial timelike scalar field profile. In both cases, the deviation from the Schwarzschild solution in GR can be characterized by a single parameter, which affects the QNFs. In the case of the stealth Schwarzschild solution, we have derived the analytical formula~\eqref{eq:QNM_const_3_rescaled} for the QNFs, showing that the QNFs are given by a simple scaling of those in GR. We have cross-checked the formula via other (semi-)analytical/numerical methods, including the so-called WKB approximation and the direct integration method (see Figure~\ref{fig:Comparison_anal_Direct} and Table~\ref{Table:QNMs_const_alpha3}). In the case of the Hayward solution, such a general analytic formula for the QNFs is not available, so that we have applied the direct integration method to find the QNFs numerically. As a complementary approach, we have also applied the Leaver's continued fraction method and the WKB approximation to obtain the QNFs perturbatively with respect to the deviation from the Schwarzschild solution in GR, which show a good agreement with the numerical results at least within a certain range of the parameter (see Figure~\ref{fig:QNM_Hayward} and Table~\ref{Table:QNMs_Hayward}).

There are several possible future directions. First, in modified gravity theories, the way how matter fields excite QNMs could be different from that in GR. Therefore, it would be intriguing to perform the time-domain analysis in the presence of the matter source to see whether the excited QNMs can be really described by the QNFs in our EFT. Second, it is definitely interesting to study dynamics of even-parity perturbations, which is much more complicated than the analysis of odd-parity perturbations. Indeed, the isospectrality between the odd- and even-parity QNM spectra is generally expected to no longer hold in the presence of modification of gravity. In addition to the QNMs, our EFT can be applied to study, e.g., the tidal Love numbers. Since the tidal Love numbers of a black hole are shown to vanish in GR~\cite{Binnington:2009bb,Hui:2020xxx,LeTiec:2020bos,LeTiec:2020spy,Chia:2020yla,Charalambous:2021mea}, non-vanishing Love numbers could serve as a smoking gun of modified gravity. Finally, it would be nice to generalize our EFT in order to incorporate the matter sector, which allows us to study, e.g., the tidal Love numbers of neutron stars. We leave these investigations to future work.

\section*{Acknowledgements}
It is a pleasure to thank Y.~Hatsuda, M.~Kimura, H.~Motohashi, K.~Nakashi, L.~Santoni, and V.~Vardanyan for useful discussions. This work was supported in part by World Premier International Research Center Initiative (WPI), MEXT, Japan. The work of K.~Takahashi was supported by JSPS (Japan Society for the Promotion of Science) KAKENHI Grant Nos.~JP21J00695, JP22KJ1646, and JP23K13101. The work of K.~Tomikawa was supported by the Rikkyo University Special Fund for Research (SFR). The work of V.~Y.~was supported by JSPS KAKENHI Grant No.~JP22K20367.

\appendix

\section{Leaver's method applied to the Hayward solution}
\label{sec:Leaver}

In this appendix, we briefly explain how Leaver's method~\cite{Leaver:1985ax} can be applied to the Hayward solution to compute the QNF. (See Subsection~\ref{sec:Hayward} for our setup.) We assume that the solution for the master equation~\eqref{eq:Q_master} can be written in the form of 
    \begin{align}
    Q(r)=e^{i\omega r_*}f^\nu\sum_{n=0}^{\infty}b_nf^n\;,
    \end{align}
with $f(r)=1-r_g/r$, so that the boundary condition at spatial infinity can be satisfied when the series~$\sum_{n=0}^{\infty}b_n$ converges. Here, $\nu$ is a complex constant to be determined such that the boundary condition at $r=r_g$ is satisfied. Substituting this ansatz into the master equation~\eqref{eq:Q_master} and reorganizing it as a series with respect to $f$, we obtain a set of 12-term recurrence relations for $b_n$'s, 
    \begin{align}\label{recurrence-relation_12-term}
    c^{(n)}_{0}(\omega)b_{n}+c^{(n)}_{1}(\omega)b_{n-1}+\cdots+c^{(n)}_{11}(\omega)b_{n-11}=0 \qquad
    (n\ge 0)\;,
    \end{align}
where each coefficient~$c^{(n)}_k$ is a function of $\omega$ and we have defined $b_n=0$ for $n<0$. In particular, the above recurrence relation for $n=0$ implies $c^{(0)}_0=0$ since we implicitly assume $b_0\ne 0$. This fixes the constant~$\nu$ as
    \begin{align}
    \nu=-2ir_g\omega\,\frac{(r_g^3+\sigma^3)^2}{r_g^3(r_g^3+4\sigma^3)}\;,
     \end{align}
which guarantees that the boundary condition at $r=r_g$ is satisfied. Indeed, Eq.~\eqref{eq:drdrs-Hayward} yields
    \begin{align}
    \frac{dr_*}{dr}=\frac{(r_g^3+\sigma^3)^2}{r_g^3(r_g^3+4\sigma^3)}\frac{r_g}{r-r_g}+{\cal O}\left((r-r_g)^0\right)\;,
    \end{align}
so that we have $f^\nu\sim e^{-2i\omega r_*}$, and hence $Q(r)\sim b_0 e^{-i\omega r_*}$ in the vicinity of $r=r_g$.

In the case of Schwarzschild solution in GR, the recurrence relation contains only three terms, for which the computation method for the QNF was established in \cite{Leaver:1985ax}. Fortunately, a recurrence relation containing more than three terms can be systematically reduced to a three-term recurrence relation by applying repeatedly the Gaussian elimination (see, e.g., \cite{Zhidenko:2006rs}). Therefore, we can recast \eqref{recurrence-relation_12-term} in the form
    \begin{align}
    \alpha_n(\omega)b_{n+1}+\beta_n(\omega)b_n+\gamma_n(\omega)b_{n-1}=0 \qquad
    (n\ge 0)\;,
    \label{recurrence-relation_3-term}
    \end{align}
where $b_{-1}=0$ and each coefficient is a function of $\omega$. For given $b_0$, this recurrence relation yields $b_i$ as a function of $\omega$ for all $i\ge 1$. Note that the relation~\eqref{recurrence-relation_3-term} can be equivalently written as 
    \begin{align}
    \frac{b_n}{b_{n-1}}=-\frac{\gamma_n}{\displaystyle \beta_n+\alpha_n\frac{b_{n+1}}{b_n}} \qquad
    (n\ge 1)\;.
    \end{align}
Therefore, we have an equation with an infinite continued fraction as follows:
    \begin{align}
    0=\beta_0+\alpha_0\frac{b_1}{b_0}
    =\beta_0-\frac{\alpha_0\gamma_1}{\displaystyle \beta_1+\alpha_1\frac{b_2}{b_1}}
    =\beta_0-\frac{\alpha_0\gamma_1}{\displaystyle \beta_1-\frac{\alpha_1\gamma_2}{\displaystyle \beta_2+\alpha_2\frac{b_3}{b_2}}}
    =\cdots\;.
    \label{infinite-continued-fraction}
    \end{align}

As mentioned earlier, in order for the boundary condition at spatial infinity to be satisfied, the series~$\sum_{n=0}^{\infty}b_n$ must converge. It should be noted that the series converges if and only if $\omega$ coincides with the QNF. Hence, for $\omega$ corresponding to the QNF, one is allowed to truncate the infinite continued fraction in \eqref{infinite-continued-fraction}, i.e., put $b_n=0$ for some large $n$. Then, Eq.~\eqref{infinite-continued-fraction} gives an algebraic equation for $\omega$, and one can obtain the QNF as a solution to this algebraic equation. This is the idea of Leaver's original paper~\cite{Leaver:1985ax}. However, as suggested by Nollert in \cite{Nollert:1993zz}, one can do a better approximation than simply truncating the continued fraction. By studying the large-$n$ behavior of the recurrence relations, the ratio~$b_n/b_{n-1}$ can be expanded as~\cite{Nollert:1993zz,Zhidenko:2006rs} 
    \begin{align}
    \frac{b_n}{b_{n-1}}=1+\frac{C_1(\omega)}{\sqrt{n}}+\frac{C_2(\omega)}{n}+\frac{C_3(\omega)}{n^{3/2}}+\frac{C_4(\omega)}{n^2}+\cdots\;.
    \label{ratio_large-n}
    \end{align}
One can substitute this ansatz into the recurrence relations to determine each coefficient order by order. Note that, instead of the reduced three-term recurrence relation~\eqref{recurrence-relation_3-term}, one could use the original relation~\eqref{recurrence-relation_12-term} to determine the coefficients in \eqref{ratio_large-n}~\cite{Zhidenko:2006rs}. For instance, the first few coefficients are given by\footnote{The coefficient~$C_1$ satisfies $C_1^2=-2ir_g\omega$. We chose the branch whose real part is negative, for which $b_n\to 0$ as $n\to \infty$. (Note that QNMs have ${\rm Re}\,\omega>0$ and ${\rm Im}\,\omega<0$, and hence $-\pi/4<\arg\sqrt{r_g\omega}<0$.) Once we choose the branch for $C_1$, the higher-order coefficients~$C_2, C_3, C_4, \cdots$ can be uniquely determined. Note also that $C_1$ and $C_2$ are independent of $\ell$.} 
    \begin{align}
    C_1=(-1+i)\sqrt{r_g\omega}\;, \qquad
    C_2=-\frac{3}{4}-ir_g\omega\,\frac{2r_g^3+\sigma^3}{r_g^3+\sigma^3}\;.
    \end{align}
Then, rather than just putting $b_n=0$ in \eqref{infinite-continued-fraction} for some large $n$, one could replace the ratio~$b_n/b_{n-1}$ by the expression~\eqref{ratio_large-n} to improve the approximation. In our computation (Section~\ref{sec:Hayward}), we truncate the expression~(\ref{infinite-continued-fraction}) up to $n = 50$ and we then replace $b_{50}/b_{49}$ in \eqref{infinite-continued-fraction} with the expansion~\eqref{ratio_large-n} up to ${\cal O}(n^{-2})$. This leaves us with an algebraic equation for $\omega$, which we denote by $P(\omega;\eta)=0$. Here, the parameter~$\eta\equiv \sigma^3/\mu^3$ controls the deviation from the Schwarzschild case. Note that we use $\mu$ (corresponding to twice the ADM mass) to normalize dimensionful quantities. Since $r=r_g$ is a solution to \eqref{eq:r_g-Hayward}, we have
    \begin{align}
    \frac{r_g}{\mu}=1+\eta-3\eta^2+15\eta^3-91\eta^4+\cdots\;,
    \end{align}
which is useful when we express $r_g$ in terms of $\mu$.

We are now interested in a branch of solutions to $P(\omega;\eta)=0$ such that $\omega\to \omega_{\rm GR}$ (i.e., the QNF in the Schwarzschild case) as $\eta\to 0$. Therefore, we assume that $\omega$ can be expanded as
    \begin{align}
    \omega(\eta)=\omega_{\rm GR}+\omega^{(1)}\eta+\omega^{(2)}\eta^2+\omega^{(3)}\eta^3+\omega^{(4)}\eta^4+\cdots\;,
    \end{align}
and substitute it into $P(\omega;\eta)=0$ to obtain
    \begin{align}
    \left(P_\eta+P_\omega\omega^{(1)}\right)\eta+\frac{1}{2}\left(P_{\eta\eta}+2P_{\eta\omega}\omega^{(1)}+P_{\omega\omega}\omega^{(1)2}+2P_\omega\omega^{(2)}\right)\eta^2+\cdots=0\;.
    \end{align}
Here, the function~$P$ and its derivatives (e.g., $P_\omega\equiv \partial P/\partial\omega$ and $P_\eta\equiv \partial P/\partial\eta$) are evaluated at $(\omega;\eta)=(\omega_{\rm GR};0)$ and we have omitted the zeroth-order term in $\eta$ which vanishes by definition. By requiring that the above equation is satisfied at each order in $\eta$, one can uniquely determine the expansion coefficients~$\omega^{(k)}$ in a successive manner. In principle, the above formalism can be applied to any multipole index~$\ell$ and any overtone, provided that the perturbative treatment is valid. We applied this formalism to obtain the formula~\eqref{QNM_Hayward_Leaver} for the fundamental $\ell=2$ QNF.

	{}
	\bibliographystyle{utphys}
	\bibliography{bib_v4}

\end{document}